\documentclass[12pt]{amsart}
\usepackage{graphicx}
\usepackage{geometry}
\usepackage[nodisplayskipstretch]{setspace}
\usepackage{bbold}
\usepackage{amssymb}
\usepackage{setspace}
 \usepackage{wasysym}
\usepackage[dvipsnames]{pstricks}
\usepackage{pst-node}
\usepackage{pst-slpe}\usepackage{booktabs}
\usepackage{pst-tree}
\usepackage{bbm}

\usepackage{amsmath, amsthm, amssymb}
\usepackage{amsmath}

 \usepackage[flushleft]{threeparttable}
\usepackage{hyperref}
\hypersetup{
    colorlinks,
    citecolor=blue,
    filecolor=black,
    linkcolor=blue,
}
\usepackage[marginal,multiple]{footmisc}

\allowdisplaybreaks
\usepackage[multiple]{footmisc}
\usepackage{amsfonts}
\usepackage{amsmath}
\usepackage{amssymb}
\usepackage{graphicx}
\usepackage{appendix}
\usepackage{multirow}
\usepackage{ulem}
\usepackage{setspace}
\usepackage[dvipsnames]{pstricks}
\usepackage{pst-node}
\usepackage{pst-slpe}
\usepackage[square,comma]{natbib}
\usepackage{amsthm}
\usepackage{lmodern}
\usepackage[T1]{fontenc}
\usepackage{mathtools}
\usepackage{MnSymbol}
\usepackage{amssymb}% http://ctan.org/pkg/amssymb
\usepackage{pifont}% http://ctan.org/pkg/pifont

\usepackage{natbib}
\usepackage{amsthm}
\DeclareMathOperator*{\argmin}{argmin}
\DeclareMathOperator*{\argmax}{argmax}

\newtheorem{lemma}{Lemma}[section]

\newtheorem{prop}{Proposition}[section]
\newtheorem{co}{Corollary}[section]
\newtheorem{remark}{Remark}
\theoremstyle{definition}
\newtheorem{definition}{Definition}

\newtheorem*{assumption1*}{Assumption 1'}

\begin{document}

\title[]{Risk and Monotone Comparative Statics without Independence}

\author[]{Collin Raymond$^\circledast$}
\author[]{Yangwei Song$^\dag$}
%\thanks{$^{\ast }$}
\thanks{~$^\circledast$ Cornell University, Email: \texttt{collinbraymond@gmail.com}}

\thanks{$^\dag$ University of Colorado Boulder, Email: \texttt{yangwei.song@colorado.edu}}
\thanks{January, 2026.  We'd like to thank seminar participants at Arizona State, Asia Meeting of the Econometric Society, Duke, Georgetown, Lingnan University, Midwest Theory Conference, Risk Uncertainty and Decision, SAET, Southern Economic Association, Time Uncertainties \& Strategies IX, University of Chicago, University of Navarra, University of Pennsylvania, University of Rochester, and helpful comments from Chris Chambers, David Dillenberger, Jean-Pierre Drugeon, Miles Kimball, Yusufcan Masatlioglu, Daniel Rappoport, Todd Sarver, Ed Schlee, and Bruno Strulovici.}
%JEL: D81 

\begin{abstract}
	\noindent  
	We extend well-known comparative results under expected utility to models of non-expected utility by providing novel conditions on local utility functions. We illustrate  how our results parallel, and are distinct from, existing results for monotone comparative statics under expected utility, as well as risk preferences for non-expected utility. Our conditions generalize existing results for specific preferences (including expected utility) and allow us to verify monotone comparative statics for novel environments and preferences. We apply our results to portfolio choice problems where preferences or wealth might change, as well as  precautionary savings.
\end{abstract}

\maketitle

 \newpage

 \section{Introduction}
 
Understanding how agents respond in situations involving risk and uncertainty is at the heart of many economic problems. These kinds of predictions allow researchers to use models for counterfactual thinking, and to derive intuitive tests that can be taken to data.  Of particular importance is understanding when the model possesses monotone comparative statics (MCS) --- i.e. behavior by the economic agent always responds in the same directional fashion to a directional change in parameters.  This enables predictions to be independent of the exact levels of behavior and parameters. For example, a classic problem is when wealthier individuals   invest more in riskier assets (\cite{arrow1965,pratt1964}).

%For example, having a positive precautionary savings motive implies that individuals should save more when future returns grow riskier, regardless of the current distribution of returns.

Although the formal study of comparative statics in economics dates back to \cite{samuelson1948foundations}, general results under certainty were not formalized in economics until \cite{milgrom1994} (although there were antecedents of this in work such as \cite{topkis1978}), and these were not systematically extended to  behavior under risk until \citet{athey}. \citet{athey}, extending many disparate results, provides a unified treatment that shows how, under expected utility (EU), assumptions on primitives like the distribution over outcomes and the Bernoulli utility function of the agent, can be aggregated in order to generate MCS predictions.  

Despite the tractability of the EU approach, many researchers have recognized that it suffers from a variety of unrealistic predictions, ranging from the Allais paradox (\citet{allais1979foundations}) to small stakes risk aversion (\citet{rabin-eu}).  In response, a plethora of models have been developed, ranging from models which maintain the linearity of indifference curves but allow them to be non-parallel (\citet{dekel1986}, \citet{gul1991}, \citet{hong1983generalization}) to those that relax linearity (\citet{chew1991mixture}, \citet{quiggin2012generalized}, \citet{kHoszegi2007reference}).  Despite the appeal of these models --- namely, their ability to better explain observed deviations from the predictions of EU --- these models are often less tractable.  Thus, it can be harder to see what predictions they make in many standard economic environments.  Researchers typically have to leverage new approaches for each specific model.

This paper develops a set of tools that extend well known MCS results under EU to models which violate the tenets of EU.  It  provides a unified approach for developing both intuitions as well as formal results that link the structure of the underlying primitives (such as payoffs, utilities and distributions) to predictions about the behavior of economic agents across a wide range of non-EU models (i.e. those that violate the Independence Axiom).  In particular, our results allow researchers to derive many standard MCS statements so long as they know the local utilities of the non-EU preferences (\cite{machina1982}) --- and these local utilities are widely known for all widely used non-EU models.  A key caveat of our approach is that, like for existing non-EU work on risk, preferences need to be ``smooth'' enough (e.g., Gateaux differentiable).

Our contributions are twofold. First, we provide results that generalize and extend existing MCS results for non-EU models, and allow us to provide novel MCS results for new environments (e.g, precautionary savings) and preferences (e.g., quadratic). Second, we show how MCS results for non-EU preferences relate to existing results MCS results under EU, as well as known results for non-EU risk preferences.  We highlight precisely why, and how, the MCS conditions for non-EU may, at first glance, appear surprising to people familiar with those two literatures.  This is because our conditions require us to account for the inherent non-separabilities across outcomes present in non-EU models.  

Both substantively and formally, our approach extends \cite{athey}, leveraging many of the technical tools in her paper.  Our main results show there are two natural ways to extend her well-known conditions for MCS, captured by Propositions \ref{prop-case1-w}/\ref{prop-body-ext1} and Proposition \ref{prop-case1}.  Her work made assumptions (such as log-supermodularity, or single-crossing) on primitives that consist of a Bernoulli utility function and distributions over outcomes.  Because non-EU models do not possess Bernoulli utility functions, we replace them with local utilities.  Local utility functions were introduced by \citet{machina1982}.  He demonstrated that many known results about risk aversion under EU could be extended to non-EU frameworks by looking at locally linear (in probabilities) approximations of the non-EU functional --- essentially, its directional derivatives.  He (along with extensions by \cite{cerreia2017}) showed that these approximations play the same role as Bernoulli utility functions in determining higher order risk preferences.

Moving from the world of EU to non-EU models involves two key changes to MCS results, and also requires us to develop new tools relative to the existing work on risk preferences in non-EU models. First we generalize the non-EU MCS problem sufficiently so that the parallels with the EU approach become transparent. This allows us to highlight that analogous results to those in the EU literature typically require additional assumptions on the primitive (specifically the utility function).  This should come as no surprise since non-EU models have weaker restrictions on the set of allowable behaviors.  Formally, this is because we need to ensure that the conditions on the local utility functions are preserved under integration.  Doing so requires us to build on the existing results in \citet{quah2012}, and show how to extend the conditions of \cite{athey} to non-expected utility functions.  

%Second, our conditions are interpretable as imposing not just restrictions on how, fixing a lottery, utility changes as we change the value of an outcome (which are the restriction under EU), but also how how this changes as the lottery changes.  In other words, in the absence of the Independence axiom, we need to consider both partial and total derivatives of the local utility functions; something that sets us apart from existing work on the risk preferences of non-EU models (\cite{machina1982,cerreia2017}).  

Second, we show that applying these conditions to specific non-EU functional can lead to conditions that appear very distinct from those under EU.  Because local utility functions are derivatives of a function that is non-linear in probabilities (unless preferences are EU), they depend  on both the direction as well as the location of the derivative.  MCS conditions under EU  look at the shape of the Bernoulli utility as outcomes change (since Bernoulli utilities, i.e. the locals of an EU preference, only depend on the direction, regardless of which lottery we are considering).  But under non-EU, our conditions jointly restrict how the local utilities change as both the direction and the location of the derivative that defines the locals shift. This generates an interplay not present in both \cite{athey} and \cite{machina1982} (and follow-up work).  The latter showed that risk preferences only depend on the change of local utilities with respect to the direction, not the location, of the derivative (i.e. we fix the location of that the derivative is being taken at and ask what happens as the direction changes).\footnote{In fact, \cite{machina1982,machina1989} understood that looking at only the changes to the direction, not the location, was restrictive, but did not propose a general way of addressing the issue.}

In Section \ref{sec-formal}, we introduce our base notation for working with non-EU preferences as well as MCS under risk and uncertainty. Then, Section \ref{sec-mot} provides  examples that sharpen our intuitions of why existing approaches to MCS and local utilities are not sufficient.  Building on a discussion in  \cite{machina1982}, we start by discussing how standard characterizations for decreasing absolute risk aversion under EU do not extend to the non-EU case. We then conduct a similar exercise, but now showing how the standard results for precautionary savings, \cite{kimball1991precautionary}, do not extend to non-EU settings. Specifically, we show that  naively applying the EU conditions for MCS through the lens of Machina's local utility approach leads to incorrect conclusions in both cases.  In fact, we show that this approach can be both too strong and too weak. In the case of decreasing absolute risk aversion, our condition is stronger than (i.e., implies) the naive analogue of the EU condition, though the converse does not hold. By contrast, for precautionary savings, the naive analogue of the EU condition implies our condition, but again the converse does not hold.

%In Section \ref{sec-mot} we begin with two specific examples in order to demonstrate why existing approaches are not sufficient for addressing monotone comparative statics under EU.  

We next provide an overview of the general MCS problem in Section \ref{sec-MCS} and discuss our approach to solving it (emphasizing that we focus on the situation where both actions and parameters are scalars).  We highlight in our formulation that (i) our conditions apply to local utilities, rather than Bernoulli utilities, and (ii) stronger conditions are required to generate MCS relative to \cite{athey} due to the need to preserve these conditions under \emph{multiple} stages of aggregation, and (iii) why we need to consider different subcases (e.g., whether the parameter impacts distributions and the action impacts payoffs or vice versa) while under EU they can be treated in a unified manner.  Our goal is to  approach the problem from a general enough perspective that we can draw  analogies to existing work for EU.

In Section \ref{sec-MCS-dist}, we discuss our leading case where actions can change the distribution the agent faces over states, while the exogenous parameter either changes  the monetary payoffs conditional on a state or alters the local utility function (e.g., by changing risk attitudes). Using our two insights, we provide sufficient conditions for MCS.  We focus on this environment because we believe they deliver results that have the widest applicability. We provide three sets of results.  The first, which is relevant for many applications, provides sufficient conditions under which the overall utility over actions and parameters has a derivative that is single-crossing in the parameter.  This is sufficient for MCS to hold under some ancillary assumptions on the differentiability and shape of the overall utility.  Our second result derives conditions for utility over the action-parameter pairs to be single-crossing 2. This result is a natural analogue to those in \cite{athey} Section III; but with the additional considerations noted above. They need to account for the fact that local utilities capture both the direction of a derivative and the location of the derivative, and they also have to account for the fact that there is an additional layer of aggregation. Our last result imposes stronger assumptions on the primitives so that overall utility over actions and parameters is supermodular.   As we discuss in Section \ref{sec-con}, our techniques can be applied to other environments and conditions.

Having established general results,  we apply them to canonical economic settings, including investment in risky assets and precautionary savings in Section \ref{sec-apps}.  We demonstrate three key insights, each through a specific example. First, we show that our approach is general enough to nest the standard comparative static considered in \citet{machina1982}: how does investment in a risky asset change when preferences (specifically, risk aversion) change?  Applying our general framework yields the results of \citet{machina1982}, which in turn imply the EU results. Our second example demonstrates that our toolkit allows us to go beyond the standard results in the literature. Revisiting an example from Section \ref{sec-mot}, we provide results on the impact of wealth on risk aversion --- a comparative static that   \citet{machina1982} explicitly noted his approach could not address. Using our MCS we formulate the appropriate notion of wealth-dependent risk-aversion and show that it suffices to imply that a decision-maker increases their investment in a risky asset as wealth increases.  Finally, we apply our approach to the other motivating example from Section \ref{sec-mot}: precautionary savings. This example has no precedent in the existing general results. We show how our conditions allow us to derive existing results for both EU and specific non-EU preferences (e.g., rank-dependent preferences and the preferences of \citet{kp}).

We conclude in Section \ref{sec-con} discussing extensions which appear in the online appendix, while Appendices \ref{sec-lemma} and \ref{sec-body-proofs} collect proofs and additional formal results for the main text of the paper.  

In the online appendix, Section \ref{sec-MCS-u}  considers the mirror image of Section \ref{sec-MCS-dist}: actions affect the utility of agents, while parameters affect  the distribution over states (which nests the approach of \cite{machina1989}). We provide general results as well as applications. Section \ref{sec-ambiguity} discusses how to extend our results to environments with subjective uncertainty --- specifically, models of ambiguity aversion.  We present results for specific models of ambiguity aversion and applications. 
%Section \ref{sec-quasi} examines conditions under which the objective function is quasiconcave in the action, an important ancillary condition we rely on for several propositions. 

\section{Literature Review}
There is a vast literature on MCS both without uncertainty and with uncertainty under EU.  The key reference points for our work (as for many others) are the seminal papers of \cite{milgrom1994}, \cite{athey}, \cite{quah2009} and  \cite{quah2012}.  Of course, our relative contribution is to extend the analysis beyond EU.

Similarly, there is an enormous literature on non-EU preferences, but a much smaller literature on comparative statics outside the EU framework.  The  existing work typically focuses  on specific economic environments and specific functional forms. \cite{schlee1994preservation} shows how standard predictions of comparative statics from EU do not carry over to non-EU settings. \citet{ormiston1999comparative} and \citet{schlee1995comparative} derive specific types of comparative statics for several non-EU functional forms.  \citet{quiggin1991comparative} provides comparative statics for a specific class of problems under the assumption of rank-dependent utility, as do \citet{haliassos2001non} and \citet{chateauneuf2016precautionary}.

\citet{machina1989} derives more general comparative statics for Frechet differentiable functions under a strong assumption: that shifts in actions do not affect the distribution of outcomes, nor the monetary value of the outcomes, but only the local utility functions directly.  This achieves what he terms as ``functional
separation of the probability distribution from the control variable''.  Under this restriction, he can leverage tools using the partial derivatives of local utility functions to derive comparative statics. \citet{hong1992differentiability} extend \citet{machina1989} to Gateaux differentiable functions. \citet{neilson1995comparative} discusses how comparative statics problems without EU have both linear and non-linear effects.  He discusses how \citet{machina1989}'s analysis focuses on ``linear'' effects from changes in parameters and actions, and shows that for some restrictive classes of preferences (e.g., betweenness preferences) the non-linear terms disappear.  Of course, these types of analyses are restrictive both in the range of problems they can address, as well as the kind of non-EU preferences to which they apply.  Our goal is to explicitly account for these non-linear terms and show how one can still  obtain tractable comparative statics. Thus, our results are more general, as we do not require this functional separation.

Using distinct tools, \citet{quah2024} establish sufficient conditions for MCS to hold when agents' preferences are represented by maxmin preferences (\cite{gs}) as well as variational preferences (\cite{massimovar}). We complement their analysis by deriving sufficient conditions that work for a broader class of ambiguity averse preferences, including the smooth ambiguity preferences of \citet{kmm}.  In the context of portfolio choice under ambiguity, \citet{gollier2011} analyzes how the demand for an uncertain asset varies with the degree of ambiguity aversion within the smooth ambiguity framework. \citet{gollier2015} examine both the smooth ambiguity and $\alpha$-maxmin models, providing conditions under which the demand for the risky asset increases with wealth. We can derive their results for the smooth ambiguity model within our general framework. \citet{cerreia2022wealth} investigate how wealth  absolute ambiguity attitudes affect portfolio choices across various ambiguity models. Since they rely on different frameworks and assumptions, their results are not directly comparable with ours.

\section{Framework} \label{sec-formal} 

\subsection{Definitions: Risk Preferences}\label{sec-differentiability}
We first introduce some technical definitions that undergird our approach.  We draw directly on the tradition of non-EU analysis begun by \cite{machina1982}.  Underlying our approach is a tradition in economic analysis that derives results for non-EU functionals that are still sufficiently ``smooth''.

%Informally, Frechet differentiable functions are a set of preferences (which nest expected utility) that are ``smooth'' enough, in a way that we formalize later.  \cite{machina1982} shows that this allows for a generalized form of expected utility analysis, which we will exploit and extend later in the paper.  

We first introduce our domain of preferences. Let $\Delta (Z)$ be the set of all cumulative distribution functions (cdf) $F$ over a compact interval $Z=[\underline{z},\overline{z}]$. Assume that the individual's preference ranking over $\Delta(Z)$ is complete, transitive, and representable by a preference functional $V : \Delta(Z) \to \mathbb{R}$.

For interpretation, suppose that elements of $Z$ are monetary outcomes (although this is not necessary), so we focus on monetary lotteries. Throughout this paper, all distributions under consideration will be from the set $\Delta (Z)$ unless otherwise specified. Impose the topology of weak convergence on  $\Delta (Z)$, which defines a sequence $\{F_n\}$ as converging to $F$ if and only if $F_n(z)\rightarrow F(z)$ at each continuity point $z$ of $F$. This topology is the weakest topology for which the expected utility functional $\int u^*(z)dF(z)$ is continuous for all continuous $u^*$.

Given two  cdfs $F$ and $G$, $F$ first-order stochastically  dominates (FOSD) $G$, denoted by $F\succsim_{FOSD} G$, if  $F(z)\leq G(z)$ for all $z\in Z$.  Throughout this paper, the order on the set of distributions $\Delta (Z)$ is $\succsim_{FOSD}$ unless otherwise specified. 

\begin{definition}\label{frechet}
The function $V: \Delta (Z)\rightarrow \mathbb{R}$ is said to be \textbf{Frechet differentiable} if there is a continuous function $u(\cdot,F): Z\rightarrow\mathbb{R}$ corresponding to each $F\in \Delta (Z)$ such that for every $G\in \Delta (Z)$,
\begin{align*}
V(G)-V(F)=\int u(z,F) d(G-F)+o(\parallel G-F\parallel), 
\end{align*}
where $o(\cdot)$ denotes a function which is zero at zero and of a higher order than its argument. 
\end{definition}

This shows a differential movement from $F$ to $G$ changes the value of $V$ by $\int u(z,F) d(G-F)$, that is, by the difference in the expected value of $u(z,F)$ with respect to $F$ and $G$. In other words, in ranking differential shifts from an initial distribution $F$, the individual acts precisely as would an EU maximizer, with \textbf{local utility function} $u(z,F)$. Throughout the paper, we assume that $u(z,F)$ is sufficiently smooth to ensure the existence of the requisite derivatives.

We next introduce two weaker notions of differentiability of $V$. 

\begin{definition}\label{def-Hadamard}
The function $V: \Delta (Z)\rightarrow \mathbb{R}$ is said to be \textbf{Hadamard differentiable}  if there is a continuous function $u(\cdot,F): Z\rightarrow \mathbb{R}$ corresponding to each $F\in \Delta (Z)$ such that for every smooth path $G(\cdot,t)$ with $G(\cdot,0)=F$,
\begin{align*}
V(G(\cdot,t))-V(F)=t\int u(z,F)dG_t(z,0)+o(t),
\end{align*}
where $G_t(z,t)$ is the derivative w.r.t. $t$.
\end{definition}
\begin{definition}\label{def-Gateaux}
The function $V: \Delta (Z)\rightarrow \mathbb{R}$ is said to be \textbf{Gateaux differentiable}  if there is a continuous function $u(\cdot,F): Z\rightarrow\mathbb{R}$ corresponding to each $F\in \Delta (Z)$ such that for every $G\in \Delta (Z)$,
\begin{align*}
V((1-t)F+t G)-V(F)=t \int u(z,F) d(G-F)+o(t). 
\end{align*}
\end{definition}
Hadamard differentiability is equivalent to Frechet differentiability in finite-dimensional spaces and is weaker in infinite-dimensional spaces. It is stronger than Gateaux differentiability. It is the weakest definition of a differential with the chain rule property (see e.g. \citet[p. 124]{nashed}). The local utility functions $u(\cdot,F)$ coincide under Frechet, Hadamard, and Gateaux differentiability whenever they exist.   Although most preferences used in the literature satisfy Hadamard differentiability, not all do.  For example, not all betweenness preferences are Hadamard, or even Gateaux, differentiable.

The expected utility functional $V(F)=\int u^*(z)dF(z)$, where $u^*$ is the Bernoulli utility function, is Frechet differentiable with the local utility function $u(z,F)=u^*(z)$ for all $F\in \Delta (Z)$.

Another class of non-EU preferences is the \textbf{rank dependent  utility (RDU)} (\citet{quiggin1982}, \citet{yaari1987}) with the functional form
\begin{align}\label{rdu}
V(F)=\int u^*(z) d \omega(F(z)),
\end{align}
where $\omega:[0,1]\rightarrow [0,1]$ is continuous, strictly increasing and onto and the utility function $u^*$ is continuous and strictly increasing. \citet{chew1987} show that RDU is not Frechet differentiable and \citet{hong1992differentiability} show that RDU is Hadamard differentiable if $\omega$ is differentiable and the derivative $\omega_1$ is bounded.\footnote{See \citet{hong1992differentiability} for some other classes of non-EU preferences and the associated conditions for Hadamard and Gateaux differentiability. 
} The corresponding local utility function is given by 
\begin{align}\label{rdu-local}
u(z,F)=\int_{\underline{z}}^z \omega_1(F(t))du^*(t). 
\end{align}

Another class of non-EU preferences arises from the induced preferences 
studied by \citet{machina1984}. He shows that even when an agent is an EU maximizer, 
her choices over temporal risky prospects need not themselves be representable as 
maximizing expected utility. To see this, consider the induced preference 
functional
\[
    V(F) := \max_{a \in A} \int \varphi(z,a)\, dF(z),
\]
where $A$ is a compact set. Note that the preferences represented by \(V\) are nonlinear in probabilities. \citet[Theorem 5]{machina1984} shows that, under quite general conditions, the induced preference functional $V$ is Frechet differentiable with local utility $u(z, F) = \varphi\bigl(z, a^{*}(F)\bigr)$, where $a^{*}(F)$ is the unique maximizer.\footnote{Dynamic mixture-averse preferences in \citet{sarver2018} also admit an induced preference representation for second-period preferences. Hence, whenever
this induced functional is differentiable, the analysis developed in our paper applies
directly.}

 \subsection{Definitions: Comparative Statics}
We next introduce some concepts from lattice theory in order to speak to modern notions of comparative statics. Given a partially ordered set $(X,\geq)$, the operations ``meet'' ($\wedge$) and ``join'' ($\vee$) are defined as follows: $x\wedge  y:= \sup \{z|z\leq x,z\leq y\}$ and $x\vee y:= \inf\{z|z\geq x,z\geq y\}$. Given a set $X\subseteq \mathbb{R}^n$ with the usual product order, operations meet and join are
\begin{align*}
x\wedge  y= (\min (x_1, y_1),..., \min (x_n, y_n))\quad\text{~~~~ and ~~~~}\quad x\vee y= (\max (x_1, y_1),..., \max (x_n, y_n)). 
\end{align*}
A partially ordered set $(X,\geq)$ is a  \textbf{lattice} if the set is closed under meet and join.

\begin{definition}
Let $(X,\geq)$ be a lattice. A function $h: X\rightarrow \mathbb{R}$ is \textbf{supermodular} if $h(x\vee y)+h(x\wedge y)\geq h(x)+h(y)$ for all $x,y\in X$. The function $h$ is \textbf{log-supermodular (log-spm)} if it is nonnegative and $h(x\vee y) h(x\wedge y)\geq h(x) h(y)$ for all $x,y\in X$.
\end{definition}

We interpret these two definitions for $X\subseteq \mathbb{R}^2$. For any $\hat{x}_1>x_1$, the supermodularity of $h$ requires that the \textit{incremental} returns to increasing the first argument, defined by $h(\hat{x}_1,x_2)-h(x_1,x_2)$, increase when the second argument $x_2$ increases; The log-spm of a positive function $h$ requires that  the \textit{relative} returns to increasing the first argument, defined by $\frac{h(\hat{x}_1,x_2)}{h(x_1,x_2)}$, increase when the second argument $x_2$ increases.

We next introduce a condition that is weaker than supermodularity and log-spm. 
\begin{definition}
Let $(S, \geq)$ be a partially ordered set. A function $g:S\rightarrow \mathbb{R}$ satisfies \textbf{single crossing (SC1)} in $s$ if for all $\hat{s}>s$,
\begin{align*}
&g(s)\geq (>)~ 0 \quad\Rightarrow\quad g(\hat{s})\geq  (>)~ 0 . 
\end{align*}
\end{definition}
The definition of SC1 says that as $s$ increases, $g(s)$ crosses zero at most once and from below.

\begin{definition}
Let $(S, \geq)$ and $(X, \geq)$ be two partially ordered sets. A function $h:X\times S\rightarrow \mathbb{R}$ satisfies \textbf{single crossing in two variables (SC2)} in $(x;s)$ if for all $\hat{x}>x$, $g(s):= h(\hat{x},s)-h(x,s)$ satisfies SC1 in $s$. 
\end{definition}

The next condition, \textit{interval dominance order}, introduced by \citet{quah2009}, is weaker than single crossing.

\begin{definition}
Let $S,X\subseteq \mathbb{R}$. A function $h:X\times S\rightarrow \mathbb{R}$ satisfies \textbf{interval dominance order} in $(x;s)$ if  
\begin{align*}
h(\hat{x},s)-h(x,s)\geq (>)~ 0 \quad \Rightarrow\quad  h(\hat{x},\hat{s})-h(x,\hat{s})\geq (>) ~0
\end{align*}
holds for all $\hat{x}$ and $x$ such that $\hat{x}>x$ and $h(\hat{x},s)\geq h(x',s)$ for all $x'\in X$ and $ x\leq x'\leq \hat{x}$, and all $\hat{s}>s$.
\end{definition}

Finally, we specify an order over sets. 

\begin{definition}
For two sets $A,B\subseteq \mathbb{R}$, we say that $A\leq B$ \textbf{in the strong set order} if for any $a\in A$ and $b\in B$ such that $a\geq b$, we  also have $b\in A$ and $a\in B$. 
\end{definition}
\section{Motivating Examples}\label{sec-mot}
With our formalities in hand, we now provide two examples demonstrating how simply modifying existing methods developed for MCS under EU, in a way directly analogous to the ``local'' utility analysis pioneered by \cite{machina1982}, will not work. We consider these examples illustrative, as both \citet{machina1982} and \citet{neilson1995comparative} point out that the analysis developed in 
\citet{machina1982} does not generally extend to MCS problems. That said, 
to our knowledge, the specific issues relating to precautionary savings 
have not been previously discussed in the literature.

Recall that \cite{machina1982}, along with a recent extension by \cite{cerreia2017}, demonstrated that one can use  local utility analysis in order to apply traditional notions of risk preferences (including higher order ones) to non-EU models.  In particular, as he and subsequent papers showed, one can look at the Gateaux derivative of a non-EU functional, i.e. the local utilities, and ask whether those local utilities possess properties that are analogous to the ones we look for in a Bernoulli utility. Specifically, he showed that one could impose conditions on the partial derivatives of the local utility function with respect to its first argument that are equivalent to the usual conditions imposed on the derivatives of the Bernoulli utility. Thus, for risk aversion, one looks for whether the local utility functionals are concave; for prudence, one looks for whether the marginal of local utilities are convex, and so on.  

This would suggest that in order to do MCS analysis, we should look at the conditions we impose on the Bernoulli utility function under EU, and impose those same conditions on the partials of local utility functions for non-EU preferences.  

Unfortunately, this would be incorrect. Our examples demonstrate that  the analogous conditions on local utility functions that are imposed on Bernoulli utility functions are neither necessary nor sufficient for the desired comparative statics to hold. 
Our first example provides a situation where the standard EU condition is implied by (but does not imply) the MCS, while the second shows the opposite. This highlights the fact that we need to develop a new toolkit to analyze non-EU setting, not just weaken or strengthen existing conditions.

 \subsection{Decreasing Absolute Risk Aversion}\label{sec-dara}
Our first example draws on an insight from \cite{machina1982}.  Consider the standard portfolio choice problem. An agent with wealth $w$  must decide how to invest it. She has two choices: a safe asset that pays a return $r>0$ and a risky asset that pays a nonnegative random return $s$. If she purchases $x$ of the risky asset and invests the remaining $w-x$ in the safe asset, she will end up with $xs + (w-x)r$. 

We want to understand when it is the case that as wealth increases, the optimal investment in the risky asset grows.  Recall that under EU, this question amounts to asking whether the Arrow-Pratt measure of risk aversion $-\frac{u_{11}^*(z)}{u_{1}^*(z)}$ is decreasing in $z$, where $u^*$ is the Bernoulli utility function --- in other words, whether agents exhibit decreasing absolute risk aversion (DARA).

To simplify the example, assume that the agent's preference functionals are smooth, in the sense that they are Frechet differentiable  in lotteries, and the objective function is  quasiconcave and differentiable in the choice $x$. \citet[p.299]{machina1982} calls such agents \textbf{diversifiers}.

As highlighted previously, given local utility function $u(z,F)$, \cite{machina1982} shows that $-\frac{u_{11}(z,F)}{u_1(z,F)}$ is the right measure of risk aversion for a non-EU functional; in that it characterizes aversion to mean preserving spreads.  Thus, the obvious analog under non-EU  for DARA is that $-\frac{u_{11}(z,F)}{u_1(z,F)}$ is decreasing in $z$. In fact, \cite{machina1982} and \cite{machina1989} note that this measure \emph{does not} allow us to answer the questions as to whether the optimal investment $x$ is increasing in wealth $w$. 

We elucidate the gap via the use of an example which leverages a specific non-EU function --- quadratic utility. \cite{machina1982} discusses a particular type of quadratic preferences:  
\begin{align}\label{quadratic}
V(F)=\int R(z)dF(z)+\frac{1}{2} [\int S(z)dF(z)]^2,
\end{align}
where $R$ and $S$ are absolutely continuous.\footnote{Quadratic preferences are studied, for example, in  \citet{machina1982},  \citet{chew1991mixture}, and \citet{Fudenberg2022}.} Consider the following parameterization: 
\begin{align}\label{quadratic-p}
R(z)=\int_{0}^{z} e^{-\alpha s}ds\quad\text{and}\quad S(z)=\int_{0}^{z} e^{-\beta s}ds,
\end{align}
where $z\in [0,\overline{z}]$ and  $\alpha, \beta\in \mathbb{R}_{++}$. \citet{machina1982} shows $V$ is Frechet differentiable with local utility function
\begin{align*}
u(z,F)=R(z)+S(z)[\int S(t)dF(t)]=R(z)+S(z) \mathbb{E}_F S(t). 
\end{align*}
Given that $-\frac{R''(z)}{R'(z)}=\alpha$ and $-\frac{S''(z)}{S'(z)}=\beta$, straightforward calculation yields that $-\frac{u_{11}(z,F)}{u_{1}(z,F)}$ decreases in $z$ for any $\alpha$ and $\beta$.

However, it is not the case that for any $\alpha$ and $\beta$ that an agent with these preferences will increase $x$ in response to an increase in $w$. In fact, we show in Section \ref{sec-ln} that this MCS will hold if $\alpha \geq \beta$. Thus, the actual condition for MCS implies the straightforward extension of the standard EU condition (i.e. whenever MCS holds, we have $-\frac{u_{11}(z,F)}{u_{1}(z,F)}$ decreases in $z$), but not the other way around. 

%The objective function in this example is concave and hence quasiconcave in $x$.

\subsection{Precautionary Savings}\label{sec-motivating} Our second example considers precautionary savings.  Consider a consumer solving the following problem: 
\begin{align*}
\max_{y\in [0,w]} V(\delta_{w-y})+ V(F_{y+\tilde{\varepsilon}}),
\end{align*}
where $w$ is the initial wealth, $y$ is saving, $\delta_{w-y}$ is the degenerate distribution that assigns unit mass to the point $w-y$, $\tilde{\varepsilon}$ is a zero-mean risk, and $F_{y+\tilde{\varepsilon}}$ is the distribution of the random variable $y+\tilde{\varepsilon}$. An individual exhibits a positive precautionary saving motive if the introduction of the additional risk  $\tilde{\varepsilon}$ to second-period income leads to an increase in optimal saving $y$.

\citet{precautionary} show that, under EU and standard smoothness assumptions, a positive third derivative of the Bernoulli utility function, $u^{*}_{111}\!\ge 0$ (together with $u^{*}_{11}\!\le 0$), is equivalent to a positive precautionary saving motive.

Recall that a preference functional $V$ exhibits \emph{downside risk aversion} if, whenever lottery $F$ third-order stochastically dominates lottery $G$, we have $V(F) \ge V(G)$. It is well known that, under EU and sufficient differentiability, prudence --- that is, $u^*_{111} \ge 0$ --- is equivalent to downside risk aversion (\cite{menezes1980increasing}). Hence, $V$ exhibits downside risk aversion if and only if it exhibits a positive precautionary savings motive. Thus, one might naturally expect a simple test for a precautionary saving motive is to examine prudence even in non-EU settings; that is, to check whether $u_{111}(z,F) \geq 0$.

This is not correct.  We provide an example using RDU \eqref{rdu}, and combine two existing results in the literature. \citet{cerreia2017} show that an RDU functional exhibits downside risk aversion if and only if it is EU; in other words, under RDU, this is true only if the probability transformation function $\omega$ is the identity function, which reduces to EU (\citet[Corollary 1]{cerreia2017}). 

However, it can also be easily demonstrated that under relatively mild conditions, RDU can exhibit positive precautionary saving motives.  For example, \cite{chateauneuf2016precautionary} prove that so long as the  utility over money is both concave and has a convex marginal, \emph{and} the probability transformation function $\omega$ is concave, then the agent exhibits a positive precautionary saving motive.  In other words, $\omega_{11} \leq 0 $, $u^*_{11} \leq 0$ and $u^*_{111} \geq 0$ characterize rank-dependent preferences exhibiting positive precautionary saving motives.  The following proposition summarizes.  

\begin{prop}
    Under EU, individuals exhibit a positive precautionary saving motive if and only if they are downside risk averse.  In contrast, under RDU, downside risk aversion implies a positive precautionary saving motive but not the opposite.  
\end{prop}

Thus, under RDU, checking for downside risk aversion would cause us to too frequently claim that preferences do not exhibit positive precautionary saving motives --- applying the EU-based condition does not lead to a characterization of precautionary saving motives. Instead, it is too strong: many preferences fail to satisfy downside risk aversion yet still display a precautionary savings motive.

\section{The Monotone Comparative Statics Problem}\label{sec-MCS}

With these examples in mind, we now define the monotone comparative statics problem. Let $\Theta\subseteq \mathbb{R}$ and $X\subseteq \mathbb{R}$. The agent's objective function is given by $U:X\times \Theta\rightarrow \mathbb{R}$, where $x\in X$ represents a choice variable and $\theta\in \Theta$ represents an exogenous parameter. Let 
\begin{align*}
X^*(\theta):=\argmax_{x\in X} U(x,\theta).  
\end{align*}
The \textbf{monotone comparative statics (MCS)} problem concerns conditions on primitives under which the agent's set $X^*(\theta)$ of optimal choices is nondecreasing in $\theta$ in the strong set order.

Under EU, \citet{athey} provides conditions on the distribution over states, the mapping from states to outcomes, and the utility function that are sufficient for MCS. Our goal is to derive analogous conditions on the same primitives, except that we replace the utility function with local utility functions.

There are three ways that the parameter and the action can impact the problem: (i) through changing the distribution over states; (ii) changing the utility function conditional on a lottery (e.g., through changing the curvature of local utility); (iii) changing the monetary payoff conditional on a state.\footnote{We want to note that these channels are not mutually exclusive --- it could be that we have a utility function with a parameter $\theta$, such that increasing $\theta$ is equivalent to giving an increased monetary payoff in every state of the world; in these cases our approach will generate equivalent conditions regardless of the classification.}  We assume that parameters and actions cannot have the same impact (i.e., they have to have distinct channels (i)-(iii)), and each only acts through a single channel.  

Although this may sound like a substantive restriction to only consider these kinds of problems, in fact it is not.  Many economic problems can be recast so that each variable acts through a single channel and the two variables operate on different channels through a change of variable (as we do in the applications), and so our distinct and single channel condition is therefore a normalization of the analytical formulation rather than a restriction on the underlying economics.

In the main text, we study situations where the action $x$ affects the distribution over states, i.e., (i), and the parameter $\theta$ affects the mapping from states to utility through (ii) or (iii). This corresponds to the standard approach where parameter $\theta$ changes the ``utility function'', and the agent takes an action $x$ that impacts the distribution over  outcomes.  These cases are particularly relevant for well-known MCS results both under EU and non-EU preferences. As discussed, our formal conditions will parallel those in \cite{athey} Section III.

The complementary case, in which the roles of $x$ and $\theta$ are reversed (i.e., where $\theta$ changes the distribution over  outcomes, while $x$ affects the utility), is analyzed in Appendix~\ref{sec-MCS-u}. Under EU, the results regarding MCS can typically (e.g., those in \cite{athey}) apply symmetrically to both cases, whereas under non-EU preferences this symmetry generally fails.

Formally, our analysis builds on the results of \citet{milgrom1994} and \citet{quah2009}. 
\citet{milgrom1994} show that MCS holds whenever $U$ is SC2 in $(x,\theta)$. It is straightforward to verify that both supermodularity and log-spm are stronger conditions than SC2. 

\citet{quah2009} propose a new order on functions, interval dominance order, which is strictly weaker than SC2 and nevertheless sufficient to guarantee MCS. Lemma~\ref{lm-IDO} below shows that, when $U$ is quasiconcave in $x$ and $U_1(x,\theta)$ is SC1, $U$ satisfies the interval dominance order and, consequently, MCS holds.\footnote{Quasiconcavity ensures that $U$ is either (i) monotonic in $x$ or (ii) has a set of maximizers that forms an interval $[\underline{x},\overline{x}]$, with preferences increasing in $x$ below $\underline{x}$ and decreasing above $\overline{x}$.} %We explore conditions under which $U$ is quasiconcave in Appendix~\ref{sec-quasi}. 
 
\begin{lemma}\label{lm-IDO}
If $U(x,\theta)$ is  quasiconcave and differentiable in $x$, and $U_1(x,\theta)$ is SC1 in $\theta$, then $U$ satisfies the  interval dominance order in $(x;\theta)$.
\end{lemma}

Thus, our goal is to find sufficient conditions on primitives under which $U(x,\theta)$ satisfies supermodularity,  SC2, or $U_1(x,\theta)$ is SC1 in $\theta$. In the following section, we pursue three distinct strategies.  First, we will provide sufficient conditions so that $U_1(x,\theta)$ is SC1 in $\theta$.  Second, we switch to providing conditions guaranteeing that $U$ is SC2.  Last, we provide conditions for $U$ to be supermodular.

\section{Actions Affect Distribution}\label{sec-MCS-dist}

Let $S=[\underline{s},\overline{s}]$ denote the set of states faced by the agent. The action $x$ affects the distribution over states, which is characterized by the probability density function (pdf) $h(s;x)$ and corresponding cdf $H(s;x)$. A given action $x$ and parameter $\theta$ induce a lottery over outcomes, so $U(x,\theta)$ is equal to $V$ evaluated at the induced lottery.

 When $\theta$ affects the local utility function directly without influencing the outcomes, we  have a fixed mapping from states to monetary outcomes.  Thus, slightly abusing notation, let the monetary outcome be  $s$. Let $\overline{V}(\theta, H(\cdot;x))$ be the preference functional with the local utility function $\overline{u}(z,\theta, F)$ (we use the bar over utilities to indicate that $\theta$ impacts them directly, rather than indirectly through the lottery over outcomes). The objective function in this case is $U(x,\theta)=\overline{V}(\theta, H(\cdot;x))$.

As we will demonstrate later through an application in Section \ref{sec-machina4}, this case nests the approach of \citet{machina1982}; e.g., where $\theta$ changes the risk aversion of the individual, and the action is the choice over lotteries.  Notice that a shift in the risk aversion of individuals does not change the monetary payoff conditional on a state, but rather how a monetary amount is translated into utility, exactly the case we are discussing here.

The more complicated case is where $\theta$ affects the local utility function through outcomes. Let $\varphi(s,\theta)$ be the outcome when state $s$ is realized. Then for each choice of $x$ and each parameter $\theta$, let $F(\cdot; x, \theta)$  denote the distribution over the monetary outcomes. Let $V$ denote the preference functional $V(F(\cdot; x, \theta))$ with local utility function $u(z,F)$.  The objective function is $U(x,\theta)=V(F(\cdot; x, \theta))$.  This case covers situations where $\theta$ changes the monetary payoff conditional on a state, e.g., when $\theta$ shifts the wealth level of the individual (see Section \ref{sec-ln} for an application).  

The key difference between the two cases is that when $\theta$ enters the utility function directly, it does not affect the induced lottery and hence does not appear in the lottery argument of the local utility function. This allows for simpler analysis.

We next present three sets of sufficient conditions under which MCS holds: $U_1$ is SC1 in $\theta$, $U$ is SC2 in $(x;\theta)$, and $U$ is supermodular. We start with conditions that are most readily applicable and then extend the analysis in later subsections.

\subsection{Sufficiency for $U_1$ to be SC1} \label{sec-sc1}

We begin by imposing relatively strong assumptions on the differentiability of the preference functional, which in turn allows us to require only relatively weak conditions on how local utility functions vary with lotteries. Moreover, these assumptions correspond to those naturally made in most applications, so our first result can be readily leveraged in applications in later sections. Our conditions guarantee that $U_1$ is SC1 in $\theta$. It then follows from Lemma \ref{lm-IDO} and \citet{quah2009} that, if  $U(x,\theta)$ is quasiconcave in $x$, MCS holds.

Specifically, we assume that preference functional is Hadamard differentiable and  differentiable in $x$. When $\theta$ enters local utility function directly (our first case), 
\begin{align}\label{eq-ambiguity}
U_1(x,\theta)=\frac{d\overline{V}(\theta, H(\cdot;x))}{dx}=\frac{d\overline{V}(\theta, H(\cdot;x+t))}{dt}|_{t=0} =\int \overline{u}(s,\theta, H(\cdot; x)) h_2(s;x)ds,
\end{align}
where the last equality follows from the definition of Hadamard differentiability, and $h_2(s;x)=\frac{\partial h(s;x)}{\partial x}$. Similarly, when $\theta$ enters the outcome (our second case), 
\begin{align*}
U_1(x,\theta)&=\frac{dV(F(\cdot;x,\theta))}{dx}=\frac{dV(F(\cdot;x+t,\theta))}{dt}|_{t=0} =\int u(z,F(\cdot;x,\theta))dF_x(z;x,\theta) \\
&=\int u(\varphi(s,\theta), F(\cdot;x,\theta)) h_2(s;x)ds,    
\end{align*}
where $F_x(z;x,\theta)=\frac{\partial F(z;x,\theta)}{\partial x}$ is just $G_t$ from Definition \ref{def-Hadamard} for the path $t\mapsto F(\cdot;x+t,\theta)$, and the last equality follows from the definition of $F(\cdot;x,\theta)$.

%  This allows us to provide conditions that are sufficient for MCS without assuming how the local utility functions depend on the distribution. 

%To be more precise, let $h_2(s;x)=\frac{\partial h(s;x)}{\partial x}$.  Hadamard differentiability implies
%\begin{align}
%&U_1(x,\theta) =\int \overline{u}(s,\theta, H(\cdot; x)) h_2(s;x)ds\\
%&U_1(x,\theta)=\int u(\varphi(s,\theta), F(\cdot;x,\theta)) h_2(s;x)ds,\nonumber
%\end{align}
%for the two cases discussed above respectively. 

Comparing the two expressions above shows that a single general formulation encompasses both cases:
\begin{align}\label{eq-6.2}
U_1(x,\theta)=\int v(s,\theta, x)h_2(s;x)ds, 
\end{align}
where $v(s,\theta, x)=\overline{u}(s,\theta, H(\cdot; x)) $ in the former case, and $v(s,\theta, x)=u(\varphi(s,\theta), F(\cdot;x,\theta))$  in the latter.

Under the situation where preferences are EU with Bernoulli $u^*$, this formula ends up being 
$U_1(x,\theta)=\int u^*(s,\theta)h_2(s;x)ds$. These two formulations are quite similar, other than the dependence of $v$ on $x$ through the lottery argument of the local utility function.

Given this similarity, it should come as no surprise that the next proposition presents conditions which are natural analogs to those known under EU.   The key steps of the proof are that conditions (2) and (3) ensure that the integral $\int v(s,\theta, x)h_2(s;x)ds$ is SC1 while Hadamard differentiability of $V$ guarantees that this is the correct expression to evaluate.\footnote{Gateaux differentiability is too weak, as it only guarantees differentiability along linear paths.}  

\begin{prop}\label{prop-case1-w}
Assume \begin{enumerate}
    \item preference functionals are Hadamard differentiable;
    \item $H(s;x)$ is  SC2 in $(-x;s)$ almost everywhere (a.e.)\footnote{Say a function $g:S\rightarrow \mathbb{R}$ is SC1 a.e. if conditions in the definition of SC1 hold for almost all $(s,\hat{s})$ pairs in $S\times S$ such that $s<\hat{s}$. A function $h:X\times S\rightarrow \mathbb{R}$ is SC2 in $(x;s)$ a.e.  if for all $\hat{x}>x$, $g(s)=h(\hat{x},s)-h(x,s)$ is SC1 in $s$ a.e.  }, and $H(s;x)$ is differentiable in $x$;
    \item  $v_1\geq 0$,  $v_1(s,\theta, x)$ is log-spm in $(s,\theta)$  a.e. for all $x$.
\end{enumerate}
Then $U_1(x,\theta)$ is SC1 in $\theta$. 
\end{prop}

The easiest way to interpret these conditions is to compare them to the equivalent result under EU, where the agent has a Bernoulli utility function.  Compared to \cite{athey}, it is clear we are working with local utilities which have an additional argument (the location of the directional derivative)

\begin{remark}\label{rem:EU-1}
Suppose the agent is an EU maximizer, then the equivalent result to Proposition \ref{prop-case1-w} has the following changes to the assumptions: we do not need (1) , and (3) will not have the requirement ``for all $x$.''  This is because (i) EU is Hadamard,  (ii) the local utility function, which reduces to the Bernoulli utility function, is independent of $x$, and (iii) the conditions imply that $U$ is SC2. 
\end{remark}

We consider the Hadamard differentiability of $V$ as primarily a technical assumption. That said, recall that not all betweenness preferences are Hadamard, or even Gateaux.

Assumption (2) restricts the way in which changes in the action can impact the distribution over states.  Recall that this means that two $H$'s, induced by different $x$'s, cross at most once as a function of the state. Moreover, as \cite{athey} notes,
it is possible that increasing $x$ raises both the mean and the riskiness of the distribution, implying a mean-risk tradeoff. 

Assumption (3) has two parts. The first implies a form of monotonicity. 
For the second, when $v(s,\theta,x)=\overline{u}(s,\theta, H(\cdot;x))$, 
the assumption that $v_1(s,\theta,x)$ is log-spm in $(s,\theta)$ is equivalent to 
$-\frac{\overline{u}_{11}(s,\theta, H(\cdot;x))}{\overline{u}_1(s,\theta, H(\cdot;x))}$ 
being decreasing in $\theta$. As $\theta$ increases, the agent becomes less risk averse 
in the sense of \citet{machina1982}.

From a technical perspective, this result is an implication of our extension of \cite{quah2012}, Lemma \ref{lm-tsc}, which allows us to aggregate up properties of the local utility functions in a way analogous to \cite{athey}.

We provide applications in Section \ref{sec-apps} to better understand these assumptions.

Although conditions  (2) and (3) seem quite specific, we can modify them in a variety of ways that are analogous to conditions under EU. 

\begin{prop}\label{prop-body-ext1} 
Replace conditions (2) and (3) with either 
\begin{enumerate}
\item $h(s;x)$ is   SC2 in $(x;s)$ a.e., $h(s;x)$ is differentiable in $x$, $v\geq 0$, and $v(s,\theta,x)$ is log-spm in $(s,\theta)$  a.e. for all $x$; or 
    \item $\int_{-\infty}^s H(t;x)dt$ is SC2 in $(-x; s)$ a.e., $H(s;x)$ is differentiable in $x$, $v_{11}\leq 0$, $\int s d(H(s;\hat{x})-H(s;x))=0$, and $-v_{11}(s,\theta,x)$ is log-spm in $(s,\theta)$  a.e. for all $x$.
\end{enumerate}
Then Proposition \ref{prop-case1-w} still holds.     
\end{prop}

Under (1), as \cite{athey} notes, the condition on $h$ is stronger than FOSD but weaker than the monotone likelihood ratio, while the conditions on $v$ restrict the levels of payoffs, and log-spm indicates that higher parameters have higher relative returns to higher states.  
 
For (2), the restrictions on $v$ are equivalent to risk aversion (i.e. $v_11\leq 0$), and that changes in risk aversion with respect to the parameter are increasing in the state (i.e. $-v_11(s,\theta,x)$ is log-spm in $(s,\theta)$).  The conditions on the distribution $H$ implies that the integrals of $H$ for two different values of $X$ cross once and that they have the same mean.  

\begin{remark}\label{rem:eu-2}
As in Remark \ref{rem:EU-1}, the corresponding result under EU does not require (1) from Proposition \ref{prop-case1-w}, and we can omit from the log-spm conditions the requirement ``for all x.'' 
\end{remark}

\subsection{Sufficiency for $U$ to be SC2}\label{sec-sc2}

Note that $U_1(x,\theta)$ being SC1 is sufficient for MCS only when $U(x,\theta)$ is quasiconcave in $x$. This restriction, together with the relatively strong differentiability assumptions underlying Propositions \ref{prop-case1-w} and \ref{prop-body-ext1}, motivates us to ask whether similar results can be obtained under weaker conditions on $U$. In this section, we strengthen the conditions on the local utility functions so that $U$ is SC2. This allows us to establish MCS without imposing quasiconcavity and under weaker differentiability assumptions on $U$. Our approach identifies conditions under which $U(\hat{x},\theta)-U(x,\theta)$ is SC1 in $\theta$. In doing this, we rely on stronger aggregation properties than in the previous section, albeit ones that still leverage the approach developed by \citet{quah2012}. 

We first derive $U(\hat{x},\theta)-U(x,\theta)$. When $\theta$ affects the local utility function directly, Gateaux differentiability implies 
\begin{align}\label{U-frechet-2}
\begin{split}
U(\hat{x},\theta)-U(x,\theta)=\int_0^1\int \overline{u}(s,\theta, H_{\alpha,x,\hat{x}}(\cdot)) d(H(s;\hat{x})-H(s;x))d\alpha, 
\end{split}
\end{align}
where $H_{\alpha,x,\hat{x}}(\cdot)=(1-\alpha) H(\cdot;x)+\alpha H(\cdot;\hat{x})$ for all $\hat{x}>x$ and $\alpha\in [0,1]$. When $\theta$ affects outcomes, Gateaux differentiability implies 
\begin{align}\label{U-frechet-1}
U(\hat{x},\theta)-U(x,\theta)=&\int_0^1 \int u(\varphi(s,\theta), F_{\alpha,x,\hat{x}}(\cdot;\theta))d (H(s;\hat{x})-H(s;x)) d\alpha,  
\end{align}
where $F_{\alpha,x,\hat{x}}(\cdot;\theta)=(1-\alpha)F(\cdot;x,\theta)+\alpha F(\cdot; \hat{x},\theta)$ for all $\hat{x}>x$ and $\alpha\in [0,1]$. Then we can write 
\begin{align}\label{U-v}
U(\hat{x},\theta)-U(x,\theta)=\int_0^1\int v(s,\theta, \alpha) d(H(s;\hat{x})-H(s;x))d\alpha,
\end{align}
where in the case of \eqref{U-frechet-2}, $v(s,\theta, \alpha)= \overline{u}(s,\theta, H_{\alpha,x,\hat{x}}(\cdot)) $ and in the case of \eqref{U-frechet-1}, $v(s,\theta, \alpha)=u(\varphi(s,\theta), F_{\alpha,x,\hat{x}}(\cdot;\theta))$.\footnote{We abuse notation slightly here as the $v$ defined in this subsection has different arguments than the $v$ in the previous subsection.  However they serve the same conceptual role.  For notational simplicity, we suppress the dependence of \(v\) on \(x\) and
\(\hat{x}\). The log-spm assumptions in Proposition~\ref{prop-case1} are
understood as conditions that must hold for each such pair. }

By way of analogy, observe that the equivalent formulation under EU would be 
\begin{align}
U(\hat{x},\theta)-U(x,\theta)=\int u^*(s,\theta) d(H(s;\hat{x})-H(s;x)).
\end{align}
Notice that the Bernoulli utility function does not depend on the lottery (i.e., $\alpha$), and thus we do not need to integrate over $\alpha$.

Because our goal is to identify sufficient conditions under which $U(\hat{x},\theta)-U(x,\theta)$ is SC1 in $\theta$, we must aggregate single-crossing properties of the local utility function. Under non-EU preferences, an additional aggregation arises from the integral over $\alpha$, which captures the integration of the local utility functions over the linear path from $H(\cdot;x)$ to $H(\cdot;\hat{x})$ (or from $F(\cdot;x,\theta)$ to $F(\cdot;\hat{x},\theta)$).  Thus, MCS under Gateaux differentiable preferences resembles a multidimensional MCS problem under EU. Just as before, relative to \cite{athey} and existing work under EU, we have (i) local utilities which have an additional argument (the location of the directional derivative) compared to Bernoulli utilities, but we also now have (ii) an extra level of integration indexed by $\alpha$. Because of  (ii), we sometimes must use stronger tools than \cite{athey} for aggregating the single-crossing property, namely those in \cite{quah2012}.  We adapt their results to establish Lemma \ref{lm-tsc} which provides  sufficient conditions on $v$ and $H$ for $U(\hat{x},\theta)-U(x,\theta)$ to be SC1.

\begin{prop}\label{prop-case1}
Assume preference functionals are Gateaux differentiable. Let one of the following hold: 
\begin{enumerate}
    \item  $h(s;x)$ is  SC2 in $(x;s)$  a.e., $v\geq 0$, and $v(s,\theta, \alpha)$ is log-spm in $(s,\theta, \alpha)$  a.e. for all $\hat{x}>x$;
    \item   $H(s;x)$ is  SC2 in $(-x;s)$ a.e., $v_1\geq 0$, and $v_1(s,\theta, \alpha)$ is log-spm in $(s,\theta, \alpha)$  a.e. for all $\hat{x}>x$; 
    \item   $\int_{-\infty}^s H(t;x)dt$ is SC2  in $(-x;s)$ a.e., $v_{11}\leq 0$,  $\int s d(H(s;\hat{x})-H(s;x))=0$, and $-v_{11}(s,\theta, \alpha)$ is log-spm in $(s,\theta,\alpha)$ a.e. for all $\hat{x}>x$. 
\end{enumerate}
Then $U(x,\theta)$ is SC2 in $(x;\theta)$.\footnote{It follows from Lemma \ref{lm-tsc} that  Proposition \ref{prop-case1} continues to hold if $\alpha$ is replaced by $-\alpha$ in the assumptions. } 
\end{prop}

Our conditions here parallel those provided in the previous subsection, with two key differences.  First, as noted, we weaken the differentiability assumption on preference functionals from Hadamard to Gateaux differentiability and drop the differentiability requirement on $h$ or $H$. We compensate by strengthening the conditions on the local utility function.  Whereas previously the condition required the local utility (or its first or second derivative) to be log-spm in $(s,\theta)$ for each $\alpha$, it now must be log-spm in  $\alpha$ as well.

The interpretation of these conditions is very similar to the interpretations provided in the previous subsection, with the addition that the log-spm conditions also depend on 
$\alpha$.  Transforming these conditions to the primitive local utility functions $u(z,F)$ and $\overline{u}(z,\theta,F)$  is straightforward. For example, we can define a partial order $\succsim_x$ on the set of distributions as follows: $H(\cdot;\hat{x})\succsim_x H(\cdot;x)$ if and only if $\hat{x}\geq x$. Moreover, assume such order satisfies
\begin{align*}
(1-\alpha') H(\cdot;x)+\alpha' H(\cdot;\hat{x})\succ_x (1-\alpha) H(\cdot;x)+\alpha H(\cdot;\hat{x})\quad\forall \alpha'>\alpha, \forall \hat{x}>x. 
\end{align*}
In both cases, the conditions on how $v(s,\theta, \alpha)$ changes in $\alpha$ can be converted to  how the local utility functions $u(z,F)$ and $\overline{u}(z,\theta, F)$ change in $F$ w.r.t. $\succsim_x$.

Recall that the equivalent EU formulation omitted $\alpha$.  The next remark shows that if the conditions omit $\alpha$, then they are sufficient for EU as well.  

\begin{remark}\label{remark-sc2}
Under EU,  the induced local utility function $v$ reduces to the Bernoulli utility function, which is independent of $\alpha$. Thus, Proposition \ref{prop-case1} remains valid after omitting $\alpha$.
\end{remark}

%log-spm implies SC2. So if Assumption \ref{as-case1} (i) is strengthened so that $f(s;x)$ is log-spm, then Proposition \ref{prop-case1} continues to hold. That's why we do not need to state the result that  parallels Proposition \ref{lm-sufficient1}.  

\subsection{Sufficiency for $U$ to be Supermodular}

We now discuss a third set of sufficient conditions; ones which  induce $U$ to be supermodular.  

Specifically, recall that  
\begin{align*}
U(\hat{x},\theta)-U(x,\theta)=\int_0^1\int v(s,\theta, \alpha) d(H(s;\hat{x})-H(s;x))d\alpha.
\end{align*}
We want to ensure that this is increasing in $\theta$.  We do so by imposing supermodularity on the local utility functions. Because the sum of supermodular functions is supermodular, this property aggregates over our integrals.

    \begin{prop}\label{prop-U-spm} 
Assume 
\begin{enumerate}
\item  preference functionals are Gateaux differentiable;
  \item  $H(\cdot; \hat{x})\succsim_{FOSD} H(\cdot; x)$ for all $\hat{x}>x$; 
    \item   $v(s,\theta, \alpha)$ is supermodular in $(s,\theta)$ for  all $\hat{x}>x$ and $\alpha\in [0,1]$. 
\end{enumerate} 
Then $U(x,\theta)$ is supermodular in $(x,\theta)$. 
\end{prop}

As before, the condition simplifies under EU.  

\begin{remark}
As in Remark \ref{remark-sc2}, the corresponding result under EU is obtained by omitting   $\alpha$.
\end{remark}

We now interpret the assumptions in Proposition \ref{prop-U-spm}. To fix ideas, suppose that $v(s, \theta, \alpha)$ is increasing in $s$.  In the case where $\theta$ affects the outcome, a sufficient condition for this is that the outcome $\varphi(s,\theta)$ is increasing in state $s$, and the local utility $u(z,F)$ is increasing in outcome $z$. As a result, the agent's utility increases with the state. Assumption (2) is then the standard requirement that a higher $x$ induces a ``better'' distribution of states in the sense of FOSD. Assumption (3) states that an increase in the state is more valuable when the parameter $\theta$ is higher. Intuitively, this implies that the agent should match a better distribution of states with a higher parameter.

%In the case where $\theta$ affects the outcomes, Assumption \ref{as-spm-x} (ii) requires $v(s, \theta, \alpha)$ be supermodular in $(s,\theta)$ for all $\alpha\in [0,1]$. This condition requires the primitive local utility function $u(\varphi(s,\theta), P(\hat{x},\theta))-u(\varphi(s,\theta), P(x,\theta)) $ be increasing in $s$ for all $\theta$. Whether this assumption is satisfied or not depends on  how $u(z,P)$ depends on $P$ and how $\kappa(s,x)$ varies with $x$. A sufficient condition for Assumption \ref{as-spm-x} (ii)  to hold is  $u(\kappa(s,x), P)$ is supermodular in $(s,x)$ for all $P$ and $u(z,P(x,\theta))$ is supermodular in $(z, x)$ for all $\theta$.\footnote{It is well-known that supermodularity is not preserved under monotone transformations, so supermodularity of $\kappa$ does not ensure supermodularity of $u(\kappa(s,x), P)$. } How the distribution $P(x,\theta)$ changes with $x$ depends on the application. One example is increasing the action $x$ increases the mean as well as the riskiness of the distribution (see   Section \ref{sec-app1}). 

%We do not assume the local utility functions are continuous in $z$ and, hence, $U$ may not be differentiable in $x$ and $\theta$. That said, Assumption \ref{as-spm-x} (ii) could be seen as quite strong.  We now provide an example of a utility function that satisfies Assumption \ref{as-spm-x} (ii) in order to demonstrate its relevancy (details of this example, as well as all others, are provided in Appendix \ref{app-examples}).

\subsection{MCS and Uniqueness of Local Utilities}\label{sec-unique}
By \citet{machina1988}, the local utility function is unique up to positive affine transformations in the following sense. Let $V$ be a smooth preference functional representing a preference relation $\succsim$ on $\Delta (Z)$ and let $\{u(\cdot,F)|F\in \Delta (Z)\}$ be the set of corresponding local utility functions. Let $V^{*}$ be another preference functional with local utility functions $\{u^{*}(\cdot, F)=a(F)+b(V(F))u(\cdot,F)|F\in \Delta (Z)\}$, where  $a:\Delta (Z) \rightarrow \mathbb{R}$ and $b: \mathbb{R}\rightarrow \mathbb{R}_{{++}}$ is a positive continuous function. Then $V^{*}$ is a monotone transformation of $V$ and represents the same preference $\succsim$.  

Recall that our assumptions on the local utility function \(u\) restrict how
\(u\) depends on outcome and lottery arguments. Since the multiplicative
constant in the affine transformation above may depend on the lottery argument,
not all of our assumptions are preserved under such transformations. More
specifically, the invariance properties of our sufficient conditions depend
both on the role of \(\theta\) and on how the conditions are formulated.

When \(\theta\) does not affect the outcome and hence does not enter the lottery
argument (case~\eqref{U-frechet-2}), the conditions stated in terms of derivatives of $v$ --- namely Assumptions~(2) and~(3) of
Proposition~\ref{prop-case1} and Assumption~(3) of
Proposition~\ref{prop-U-spm} --- are preserved under such transformations. By
contrast, when \(\theta\) affects the outcome (case~\eqref{U-frechet-1}), these
same conditions need not be preserved.

When the preference functional is Hadamard differentiable, conditions
formulated in terms of derivatives of \(v\) --- such as Assumption~(3) in
Proposition~\ref{prop-case1-w} and Assumption~(2) in
Proposition~\ref{prop-body-ext1} --- are preserved under such transformations,
regardless of whether \(\theta\) affects the outcome. Finally, note that
conditions requiring log-spm of the local utility function \(v\) itself --- namely
Assumptions~(1) of Propositions~\ref{prop-body-ext1} and
\ref{prop-case1} --- are not preserved even under standard affine transformations
in the EU framework.

\section{Applications}\label{sec-apps}

We now present three applications of our results. First, we show that our approach can replicate the existing result in the literature that can be obtained by naively using the approach of \cite{machina1982} --- we provide conditions for comparative risk aversion. We then revisit the two motivating examples, for which the approach of \cite{machina1982} does not work: changes to risk-taking w.r.t. changes in wealth and precautionary savings.  The former complements \cite{machinadara}.  In all cases we discuss how our approach nests existing results (including those for EU as well as known non-EU results).

Our examples also highlight the distinction between the two cases discussed previously: Section \ref{sec-machina4} examines the setting in which $\theta$ directly affects utility, changing the mapping from lotteries to utility levels, whereas Section \ref{sec-ln} analyzes the case where $\theta$, in this application wealth, changes monetary payoffs conditional on a state.

All applications rely on Proposition~\ref{prop-case1-w} and therefore impose relatively strong differentiability conditions.  Of course, we can instead leverage Propositions \ref{prop-case1} and \ref{prop-U-spm} to obtain similar results, but with stronger conditions on the local utilities.

\subsection{Portfolio Choice and Changes to Risk Preferences}\label{sec-machina4}

We first demonstrate how our approaches nest those employed by \citet{machina1982}.  In particular, we replicate his Theorem 4, but using the language of log-spm.

Consider the same portfolio choice problem as in Section \ref{sec-dara}. Recall that the safe asset yields a constant return $r>0$, while the risky asset yields a random return $s \in [0,\overline{s}]$. Let $F_{\tilde{s}}$ denote the cdf of this random return, and assume that $\mathbb{E}[s] > r$. Let $\theta$ be a parameter which governs risk aversion. Thus, $\theta$ does not affect outcomes but affects local utility functions directly and the action $x$ affects the distribution of outcomes $z= wr + x (s-r)$.   Following the notation in Section \ref{sec-MCS-dist}, let $\overline{V}(\theta, H(\cdot;x))$ be a preference functional with local utility function $\overline{u}(z,\theta,F)$.

\begin{prop}\label{prop-asset3-frechet}
Suppose $\overline{V}$ is Hadamard differentiable and $U$ is quasiconcave in $x$. If $\overline{u}_1(z, \theta,F)$ is log-spm in $(z,\theta)$  a.e. for all $F$, then  an increase in $\theta$ leads to an increase in the optimal choice $x$.
\end{prop}

Note that by construction, the distribution of $z$ is $H(z;x)=F_{\tilde{s}}(\frac{z-wr}{x}+r)$, which is SC2 in $(-x;z)$ a.e., that is, Assumptions (2) in  Proposition \ref{prop-case1-w} is satisfied. The result is thus a direct application of Proposition \ref{prop-case1-w} and Lemma \ref{lm-IDO}.

The log-spm of $\overline{u}_1(z,\theta,F)$ implies that $-\frac{\overline{u}_{11}(z,\theta,F)}{\overline{u}_1(z,\theta,F)}$ decreases in $\theta$ for all $z$ and $F$. An alternative, but also well-known way of stating an equivalent assumption is  $\overline{u}(\cdot,\theta,F)$ is a concave transformation of $\overline{u}(\cdot,\hat{\theta},F)$ for every $F$ and $\hat{\theta}>\theta$. Consequently, an agent with a lower value of $-\frac{\overline{u}_{11}(z,\theta,F)}{\overline{u}_1(z,\theta,F)}$ invests more in the risky asset. Hence, agent $\hat{\theta}$ is less risk averse than agent $\theta$ in the sense of \citet{machina1982}, confirming that $-\frac{\overline{u}_{11}(z,\theta,F)}{\overline{u}_1(z,\theta,F)}$ is a proper measure of absolute risk aversion.

\begin{remark}
Under EU, let $u^*(z, \theta)$ denote the Bernoulli utility function.  Proposition \ref{prop-asset3-frechet} simplifies to the following: if $u^*_1(z,\theta)$ is log-spm a.e., then an increase in $\theta$ leads to an increase in the optimal choice $x$.     
\end{remark}

Proposition \ref{prop-asset3-frechet} also directly implies existing results for non-EU functionals.  Here we show that it implies the results of  \cite{chew1987} which assume the known non-EU model \eqref{rdu} of  RDU. Suppose that  the parameter $\theta$ measures risk aversion and let $\omega(\cdot,\theta)$ and $u^*(\cdot,\theta)$ be the probability transformation function and utility function of an RDU agent respectively. 

\begin{co}\label{co-chew1}
Assume RDU with $u^*_{11}(z,\theta)<0$ for all $z$ and $\theta$. If $\omega_1(p,\theta)$ and $u^*_1(z,\theta)$  are log-spm,   then  an increase in $\theta$ leads to an increase in the optimal choice $x$.
\end{co}

Note that $\omega_1$ and $u^*_1$ being log-spm is equivalent to $\omega(\cdot,\theta)$ and $u^*(\cdot,\theta)$ being concave transformations of $\omega(\cdot,\hat{\theta})$ and $u^*(\cdot,\hat{\theta})$ for all $\hat{\theta}>\theta$. Hence, Corollary \ref{co-chew1}  states that  agent $\hat{\theta}$ is less risk averse and invests more in the risky asset than agent $\theta$ for any initial wealth.

%Let $V$ and $\hat{V}$ be two RDU preference functionals with respective utility functions $v$ and $\hat{v}$ and probability transformation functions $g$ and $\hat{g}$. We are going to show the next result is an immediate implication of Proposition \ref{prop-asset3-frechet}.

\subsection{Portfolio Choice and Changes in Wealth}\label{sec-ln}
We next turn to revisiting our example from  Section \ref{sec-dara} --- how changes in wealth impact risk-taking.  Recall this is an example where the naive application of the  \cite{machina1982} approach does not work.  

Recall that the basic setup is the same as in Section \ref{sec-machina4}, but let $t=x(s-r)$. The parameter  here is the wealth $w$ that affects the outcome  $\varphi(t,w)= wr+t$ and the action $x$ affects the distribution of $t$ with distribution function $H(t;x)=F_{\tilde{s}}(\frac{t}{x}+r)$.

In order to obtain MCS with differentiable preference functionals, we introduce an assumption:  differentiable decreasing absolute risk aversion.

\begin{definition}
A differentiable preference functional \(V\) exhibits 
\textbf{differentiable decreasing absolute risk aversion (D\text{-}DARA)}  if 
\(-\frac{u_{11}(y+t, F_{y+\tilde{t}})}{u_1(y+t, F_{y+\tilde{t}})}\) is 
decreasing in \(y\) a.e., for all \(t\) and \(\tilde{t}\).\footnote{Recall that \(F_{y+\tilde{t}}\) denotes the distribution of the random variable
\(y+\tilde{t}\). Note that \(t\) is an arbitrary scalar and \(\tilde{t}\) is an
arbitrary random variable, and the two need not be related.}
\end{definition}

D-DARA is  quite permissive. In particular, it allows  $-\frac{u_{11}(z,F)}{u_1(z,F)}$ either to increase in $z$ while decreasing in $F$, or  to decrease in $z$ while increasing in $F$.\footnote{The latter pattern corresponds to \citet{machina1982}'s Hypotheses~I and~II, which he argues are consistent with a lot of experimental evidence on  systematic violations of the Independence axiom. } %More generally, MCS under non-EU preferences depends on how the local utility varies with the distribution---what \citet{machina1989} calls the “second-order properties of preferences.”
Equivalently, D-DARA says that $u_{1}(y+t, F_{y+\tilde{t}})$  is log-spm in $(t,y)$. Therefore, D-DARA implies Assumptions (3) from Proposition \ref{prop-case1-w} by taking $\theta=w$. Under EU,  D-DARA reduces to the standard DARA requirement that $-\frac{u^*_{11}(z)}{u_1^*(z)}$ falls in $z$, where  $u^*$ is the Bernoulli utility.

The next MCS result follows directly from Proposition \ref{prop-case1-w} and says that  if the preference functional exhibits D-DARA, then the agent will invest more in the risky asset the greater is her initial wealth.

\begin{prop}\label{prop-asset2-frechet}
Suppose $V$ is Hadamard differentiable and exhibits D-DARA, and $U$ is quasiconcave in $x$.  Then an increase in $w$ leads to an increase in the optimal choice of $x$. 
\end{prop}
%\footnote{Proposition \ref{prop-asset2-frechet} is consistent with the result of \citet[Proposition 3]{neilson1995comparative}.}

\begin{remark}
    Under EU, let $u^*(z, \theta)$ denote the Bernoulli utility function.  Proposition \ref{prop-asset2-frechet} simplifies to the following: if $u^*_1(z,\theta)$ satisfies DARA, then an increase in $w$ leads to an increase in the optimal choice of $x$.  
\end{remark}

%\citet{machinadara} argues that the condition $-\frac{u^*_{11}}{u_1^*}$ decreasing in wealth fails to capture decreasing risk aversion when both initial wealth and wealth increments are stochastic, as noted earlier by \citet{ross1981some}. To address this issue, 

\citet{machinadara} considered a distinct comparative static for non-EU preference: when do individuals invest more in the risky asset as wealth increases when both initial wealth and wealth increments are stochastic (building on the work of \citet{ross1981some}).  He defines \emph{generalized DARA} under Frechet differentiability as the property that  $-\frac{u_{11}(z,F)}{\int u_1(t,F)dF(t)}$ decreases in both $z$ and $F$. The generalized DARA and D-DARA conditions are independent: neither implies the other. In particular, generalized DARA cannot hold under EU with a concave Bernoulli utility function, whereas D-DARA reduces to the standard DARA condition under EU. However, \citet{machinadara} shows that generalized DARA can also imply our MCS result.

%To see D-DARA does not imply the generalized DARA, note that under EU,  D-DARA can be satisfied but  the generalized DARA cannot.  To see the generalized DARA  does not imply  D-DARA, note that $-\frac{u_{11}(z,F)}{\int u_1(t,F)dF(t)}$ decreasing in $z$ is equivalent to $u_{111}\geq 0$. However, $u_{111}\geq 0$ does not necessarily imply D-DARA since whether D-DARA is satisfied depends also on how $u$ depends on the second argument.  

%\citet{machinadara} gives a definition of generalized DARA under Frechet differetiability: $-\frac{u_{11}(z,F)}{\int u_1(t,F)dF(t)}$ decreases in $z$ and $F$. Strong D-DARA and the generalized DARA defined in \citet{machinadara} do not imply each other. To see Strong D-DARA does not imply the generalized DARA, note that under EU,  Strong D-DARA can be satisfied but  the generalized DARA cannot be satisfied given the assumption that $u$ is concave. To see the generalized DARA  does not imply Strong D-DARA, note that $-\frac{u_{11}(z,F)}{\int u_1(t,F)dF(t)}$ decreasing in $z$ is equivalent to $u_{111}\geq 0$. However, $u_{111}\geq 0$ does not necessarily imply $-\frac{u_{11}(z,F)}{u_1(z,F)}$ decreasing in  $z$. On the other hand, $-\frac{u_{11}(z,F)}{\int u_1(t,F)dF(t)}$ decreases in $F$ if and only if $\int\frac{ u_1(t,F)}{-u_{11}(z,F)}dF(t)$ increases in $F$. If $u_{11}<0$, this implies $-\frac{u_1(t,F)}{u_{11}(z,F)}$ increases in $F$, which further implies our condition that $-\frac{u_{11}(z,F)}{u_1(t,F)}$ decreases in $F$. 

We next derive the D-DARA conditions for three specific non-EU preferences.  First we consider RDU, and show that  Proposition \ref{prop-asset2-frechet} implies Theorem 2 in  \citet{chew1987}.

\begin{co}
Assume RDU with $u^*_{11}(z,\theta)<0$ for all $z$ and $\theta$. If $-\frac{u^*_{11}(z)}{u^*_1(z)}$ is decreasing in $z$, then an increase in wealth $w$ leads to an increase in the optimal investment in the risky asset.
\end{co}

Given Proposition \ref{prop-asset2-frechet}, the proof is straightforward.  Observe from our definition of the local utility function of RDU,  \eqref{rdu-local}, that for $y$, $t\in [\underline{t},\overline{t}]$ and $F_{\tilde{t}}\in \Delta [\underline{t},\overline{t}]$, 
\begin{align}
u(y+t,F_{y+\tilde{t}})=\int^t_{\underline{t}} \omega_1(F_{\tilde{t}}(s))du^*(y+s). 
\end{align}
This implies 
\begin{align*}
u_1(y+t,F_{y+\tilde{t}})=\omega_1(F_{\tilde{t}}(t))u^*_1(y+t)\quad\text{a.e.} 
\end{align*}
It is then easily seen that  D-DARA reduces to $-\frac{u^*_{11}(z)}{u^*_1(z)}$ decreasing in $z$. That is, the  utility function $u^*$ displays the standard DARA in the sense of Arrow-Pratt. 

We can also revisit the example of quadratic preferences \eqref{quadratic} used as a motivating example.  Under the parametrizations \eqref{quadratic-p}, straightforward calculation yields that $-\frac{u_{11}(y+t,F_{y+\tilde{t}})}{u_{1}(y+t,F_{y+\tilde{t}})}= \beta+\frac{\alpha-\beta}{1+e^{(\alpha-\beta)(y+t)}\int S(y+\tau)dF_{\tilde{t}}(\tau)}$ and D-DARA is satisfied  if and only if $\alpha\geq \beta$. 
%$\alpha\geq \beta$ does not gurantee $x$ increases in $w$ since this result holds under quasiconcavity of the objective function. There exist values of $\alpha>\beta$ such that the objective function is strictly concave in $x$. In those situations, $x$ increases in $w$. 

Recently, \citet{mu2024} introduced a class of non-EU preferences that are invariant to background risk:\footnote{Preferences are invariant to background risk if the addition of an independent background risk does not change preferences.}
\begin{align*}
V(F)=\int \frac{1}{a}\ln \mathbb{E}_{F} e^{az}d\mu(a),
\end{align*}
where $\mu$ is a probability measure over the usual coefficient of absolute risk aversion.  Our conditions allow us to easily check how these individuals behave in respons to changes in  wealth: the local utility function of $V(F)$ is 
\begin{align*}
u(z,F)=\int \frac{1}{a}\frac{ e^{az}}{\mathbb{E}_{F} e^{az}} d\mu(a).
\end{align*}
Clearly, $u(y+t, F_{y+\tilde{t}})$ is constant in $y$ and, hence, $-\frac{u_{11}(y+t,F_{y+\tilde{t}})}{u_1(y+t,F_{y+\tilde{t}})}$ is constant in $y$. Thus, it is constant absolute risk aversion using our measure. 

\subsection{Precautionary savings}\label{sec-precautionary}

We now return to the second of our original motivations, and discuss precautionary savings, extending the analysis in  \citet{precautionary} to non-EU settings.

%Recall that following \citet{machina1982}, \citet{cerreia2017} show that an agent with Gateaux differentiable preferences is prudent if and only if $u_{111}(z,F) \geq 0$ for all $z$ and $F$. \cite{precautionary} shows that, under EU, $u^*_{111} \geq 0$ (along with $u^*_{11} \leq 0$)  is equivalent to a positive precautionary savings motive.  Moreover, he shows that one can order the strength of agents' precautionary savings motives by looking at $-\frac{u^*_{111} }{u^*_{11}}$.   Such equivalences no longer hold without EU.  Instead, as we show, that conditions depend on total derivatives of the local utility function.  

We will derive two results: first an interpersonal measure of the strength of the precautionary saving motive, and second,  a criterion for determining whether a given individual exhibits a positive precautionary saving motive.  The latter  allows us to directly address the example in Section \ref{sec-motivating}.

Analogously to our discussion of risk in Section \ref{sec-machina4}, 
we let $\theta$ modify the agent's attitudes. Let $\overline{V}(\theta, F)$ be the preference functional. Recall the precautionary saving problem from Section \ref{sec-motivating}:
\begin{align}\label{precautionary}
\max_y \overline{V}(\theta,\delta_{w-y})+ \overline{V}(\theta,F_{y+\tilde{\varepsilon}}),
\end{align}
where $w$ is initial wealth, $y$ is saving, and $\tilde{\varepsilon}$ is a zero-mean risk.\footnote{In the precautionary saving problem, only the second-period preferences are relevant; for notational simplicity, we assume that preferences in both periods are represented by $\overline{V}$.}  The first order condition for the optimal level of saving $y$ is 
\begin{align}\label{foc}
-\frac{\partial \overline{V}(\theta,\delta_{w-y})}{\partial y}=\frac{\partial \overline{V}(\theta,F_{y+\tilde{\varepsilon}})}{\partial y}. 
\end{align}
If the agent exhibits decreasing marginal utility of money, i.e., if $\overline{V}(\theta,F_{y+\tilde{t}})$ is concave in $y$ for all $\tilde{t}$, the risk represented by $\tilde{\varepsilon}$ will cause additional saving whenever 
\begin{align*}
\frac{\partial \overline{V}(\theta,F_{y+\tilde{\varepsilon}})}{\partial y}>\frac{\partial \overline{V}(\theta,\delta_{y})}{\partial y},
\end{align*}
that is, when the risk $\tilde{\varepsilon}$ raises the marginal utility of saving. Therefore, precautionary saving motive is related to the first order derivative of the preference functional. 

%To define comparative precautionary saving motives, we use the notion of the compensating precautionary premium, adapted from \citet{precautionary}. For all $y, \tilde{t}$, and zero-mean risk $\tilde{\varepsilon}$, define the \textbf{compensating precautionary premium} $\phi^\theta$ as follows
%\begin{align*}
%\frac{d \overline{V}(\theta, F_{y+\tilde{t} +\tilde{\varepsilon}+\phi^\theta(\tilde{\varepsilon},y+\tilde{t})})}{dy}=\frac{d \overline{V}(\theta, F_{y+\tilde{t}})}{dy}. 
%\end{align*}
%Then by definition, $\phi^\theta(\tilde{\varepsilon}, y+\tilde{t})$ can compensate for the effect of the risk $\tilde{\varepsilon}$ on second-period marginal utility so that the first period saving would be unaltered by the addition of the risk $\tilde{\varepsilon}$ plus the compensating precautionary premium $\phi^\theta(\tilde{\varepsilon}, y+\tilde{t})$. Intuitively, the compensating precautionary premium shows how much wealth is needed to compensate for the effect of the risk $\tilde{\varepsilon}$ on saving. Therefore, for a fixed a level of saving and risk, a larger compensating precautionary premium suggests stronger precautionary saving motive.  

%\begin{definition}
%We say that agent  $\theta$ exhibits a \textbf{stronger precautionary saving motive than} agent $\hat{\theta}$ if the corresponding compensating precautionary premium satisfies $\phi^\theta(\tilde{\varepsilon}, y + \tilde{t}) \geq \phi^{\hat{\theta}}(\tilde{\varepsilon}, y + \tilde{t})$ for all $y$, $\tilde{t}$, and zero-mean risks $\tilde{\varepsilon}$.
%\end{definition}

To apply our MCS results, we can follow \cite{precautionary} and consider the following maximization problem based on the first order condition for the optimal level of savings:
\begin{align}\label{objective-prudence}
\max_{x} \; -\frac{\partial \overline{V}(\theta, F_{y+x\tilde{s}})}{\partial y}, \footnotemark
\end{align}
\footnotetext{Unlike the setting studied in Section~\ref{sec-MCS-dist}, the objective function here is not a preference functional itself but the derivative of one. Nonetheless, the analysis proceeds analogously.}where $\tilde{s}$ is a risk with pdf $f_{\tilde{s}}$, cdf $F_{\tilde{s}}$, and $\mathbb{E}\tilde{s}>0$. Intuitively, the objective function captures how rapidly the marginal utility changes as risk increases. 

%Under EU, problem~\eqref{objective-prudence} reduces to
%\begin{align*}
%\max_{x} \; \mathbb{E}\big[-u^*_1(y+x\tilde{s},\theta)\big].
%\end{align*}
%\citet{precautionary} demonstrates that the agent with the weaker precautionary saving motive chooses a larger maximizer for all $y$.\footnote{The counterpart for Non-EU is proved in Proposition \ref{prop-premium} in the appendix.} This result parallels \citet{arrow1965}, who shows that a less risk-averse agent selects a larger maximizer in the problem $\max_x \mathbb{E} u^*(y+x\tilde{s})$. More generally, \citet{precautionary} establishes that any theorem about risk aversion can be translated into a result about precautionary saving by substituting $-u^*_1$ for $u^*$. Instead, we show that under Non-EU, any theorem about risk aversion can be applied to precautionary saving by substituting  $-\frac{d\overline{u}(y+t,\theta, F_{y+\tilde{t}})}{dy}$ for $\overline{u}(y+t,\theta,F_{y+\tilde{t}})$.  

\begin{prop}\label{prop-precaut-frechet}
Assume $\overline{V}$ is twice Hadamard differentiable and $U$ is  quasiconcave in $x$.  Assume
\begin{enumerate}
    \item $\frac{d\overline{u}(y+t,\theta, F_{y+\tilde{t}})}{dy}$ is strictly decreasing in $t$ for all $y$, $\tilde{t}$, and $\theta$;
    \item $-\frac{d\overline{u}_1(y+t,\theta, F_{y+\tilde{t}})}{dy}$ is log-spm in $(t,\theta)$ a.e. for all $y$ and $\tilde{t}$.
\end{enumerate}
Then an increase in $\theta$ leads to an increase in the optimal choice of $x$.\footnote{We want to mention that the conclusion in Section \ref{sec-unique} that Assumption (3) in Proposition \ref{prop-case1-w} is preserved under equivalent affine transformations does not apply here.  The reason is the objective function \eqref{objective-prudence} in this application is the  derivative of the preference functional, rather than the preference functional itself. A simple monotone transformation of the  preference functional thus does not yield a monotone transformation of the objective function. Intuitively, this is because higher-order risk preferences are not preserved by monotone transformations and, hence, local utility functions $u(\cdot,F)$ and $a(F)+b(V(F))u(\cdot,F)$ are not equivalent in terms of  precautionary saving motives.}
\end{prop}

\begin{remark}
  Under EU,  let $u^*(z,\theta)$ denote the Bernoulli utility function.  Proposition \ref{prop-precaut-frechet} simplifies to the following result (\cite{precautionary}).  If  $u^*_1(z, \theta)$ is strictly decreasing in $z$ for all $\theta$ and $-u^*_{11}(z, \theta)$ is log-spm a.e. in $(z,\theta)$, then an increase in $\theta$ leads to an increase in the optimal choice of $x$.

\end{remark}

%Assumption (2) of Proposition \ref{prop-precaut-EU} says that $-\frac{u^*_{111}(z, \theta)}{u^*_{11}(z, \theta)}$ is decreasing in $\theta$. Thus, Proposition \ref{prop-precaut-EU} suggests that $-\frac{u^*_{111}(z, \theta)}{u^*_{11}(z, \theta)}$ is the appropriate measure of the strength of precautionary saving motive. 

A key conceptual insight of \citet{precautionary} was that that any theorem about risk aversion can be translated into a result about precautionary saving by substituting $-u^*_1$ for $u^*$. Our results extend this, showing that for non-EU preferences, results about risk aversion can be applied to precautionary saving by substituting  $-\frac{d\overline{u}(y+t,\theta, F_{y+\tilde{t}})}{dy}$ for $\overline{u}(y+t,\theta,F_{y+\tilde{t}})$.  

Under non-EU preferences, Assumption (1) of Proposition \ref{prop-precaut-frechet} differs from the  condition for risk aversion, $\overline{u}_{11}\leq 0$. In fact,  Lemma \ref{lem-con-new} in the appendix proves that Assumption (1) implies decreasing marginal utility of money, that is, $\frac{\partial \overline{V}(\theta, F_{y+\tilde{t}})}{\partial y}$ decreases in $y$.\footnote{These two conditions are distinct and do not imply each other. To see this, note that
\begin{align*}
\frac{d\overline{u}_1(y+t,\theta, F_{y+\tilde{t}})}{dy}=\overline{u}_{11}(y+t,\theta,F_{y+\tilde{t}})+\frac{\partial \overline{u}_1(y+t,\theta, F_{s+\tilde{t}})}{\partial s}|_{s=y}. 
\end{align*}
Obviously, $\frac{d\overline{u}_1(y+t,\theta, F_{y+\tilde{t}})}{dy}\leq 0$ does not imply $\overline{u}_{11}(y+t,\theta, F_{y+\tilde{t}})\leq 0$ and  vice versa. A sufficient condition for both risk aversion condition and Assumption  (1) is $\overline{u}_1(z,\theta, F)$ is decreasing in $z$ and $F$.}  Assumption (2) of Proposition~\ref{prop-precaut-frechet}  is equivalent to $-\frac{d \overline{u}_{11}(y+t,\theta,F_{y+\tilde{t}})/d y}{ d\overline{u}_1(y+t,\theta,F_{y+\tilde{t}})/dy}$  decreasing in $\theta$. This thus motivates a \textit{global} measure of the strength of the precautionary saving motive, defined as 
\begin{align}\label{measure-pru}
\eta^\theta_{F_{y+\tilde{t}}}(t) = -\frac{ \, d\overline{u}_{11}(y+t,\theta, F_{y+\tilde{t}})/dy \, }{ \, d\overline{u}_1(y+t,\theta, F_{y+\tilde{t}})/dy \, } \quad \forall\, y,t,\tilde{t}.
\end{align}
For any $\hat{\theta}>\theta$, Proposition~\ref{prop-precaut-frechet} implies that agent~$\theta$ exhibits a stronger precautionary saving motive than agent~$\hat{\theta}$ whenever $\eta^\theta_{F_{y+\tilde{t}}}(t)\geq  \eta^{\hat{\theta}}_{F_{y+\tilde{t}}}(t)$ for all $y,t, \tilde{t}$. 
%Standard arguments show that this condition holds if and only if,
%for each $y$ and $\tilde{t}$, there exists a strictly convex function 
%$g:\mathbb{R}\to\mathbb{R}$ such that $\frac{d\overline{u}(y+t,\theta, F_{y+\tilde{t}})}{dy} =  g\!\left( \frac{d\overline{u}(y+t,\hat{\theta}, F_{y+\tilde{t}})}{dy}
%\right)$ for all $t$. 

We are now ready to state the condition characterizing a positive precautionary saving motive.   %Lemma~\ref{lm-neutral} in the appendix shows that if  $\eta^{\hat{\theta}}_{F_{y+\tilde{t}}}(t)=0$,  then agent~$\hat{\theta}$'s optimal saving remains unchanged when her second-period income becomes riskier.  
Proposition~\ref{prop-precaut-frechet} implies that agent~$\theta$ exhibits a 
\emph{positive precautionary saving motive} whenever $\frac{d\overline{u}_{11}(y+t,\theta, F_{y+\tilde{t}})}{dy} \geq 0$. Thus, the presence of a positive precautionary saving motive is linked to the 
\emph{convexity of the total derivative} of the local utility function, rather than to the convexity of its partial derivative.

\begin{prop}\label{co-precautionary}
Suppose $V$ is Hadamard differentiable. If $\frac{d\overline{u}_1(y+t,\theta, F_{y+\tilde{t}})}{dy}<0$ and $\frac{d\overline{u}_{11}(y+t, \theta, F_{y+\tilde{t}})}{dy} \geq 0$ for all $y,t$, and $\tilde{t}$, then  agent $\theta$ exhibits a positive precautionary saving motive.
\end{prop}

\begin{remark}
  Under EU, Proposition \ref{co-precautionary} simplifies to the following result.  If  $u^*_1(z, \theta)$ is strictly decreasing in $z$ for all $\theta$ and $u^*_{111}(z,\theta)\geq 0 $ for all $z$, then agent $\theta$ exhibits a positive precautionary savings motive.
\end{remark}

Returning to the motivating example, this highlights the difference between downside risk aversion and precautionary saving in non-EU settings. Recall that 
\citet[Corollary 1]{cerreia2017}  found that a preference functional $V$ is consistent with the third-order stochastic dominance if and only if  $u_{1}(z,F)$ exists and is convex.  Notice that under differentiability, this implies we look at the third partial derivative of the local utility.  In contrast, our MCS involves a total derivative --- we need to think about the interaction between changing the direction of the derivative and its location.  

We can also use these propositions to both speak to existing results on precautionary savings, and develop new measures of comparative precautionary saving, for non-EU preferences.

First, we revisit the motivating example using the case of RDU.  Proposition \ref{co-precautionary} immediately generates the result of \citet{chateauneuf2016precautionary}.

\begin{co}\label{prop-rdu-pos}
Assume RDU with $u^*_{11}(z,\theta)<0$ for all $z$ and $\theta$. If $\omega_{11}\leq 0$ and $u^*_{111}\geq 0$, then the agent exhibits a positive precautionary saving motive. 
\end{co}

% which under RUD, implies that $g$ is the identity function (and so preferences are EU). In contrast, Corollary \ref{co-prudent} above shows that for MCS to hold, we only need  local utilities $u(z,F)$ to be differentiable a.e. and $du(y+t, F_{y+\tilde{t}})/dy$ to be convex in $t$, which can be satisfied by RDU. It is also clear from the analysis above that how properties of local utilities can be translated to properties of the utility function $u^*$ and probability transformation function $g$. 

We can also  use Proposition \ref{prop-precaut-frechet} to provide new condition that allows us to order individuals using a ``stronger precautionary saving motive'' 
\begin{co}\label{co-prudent}
Assume RDU with $u^*_{11}(z,\theta)<0$ and $u^*_{111}(z,\theta)\geq 0$ for all $z$ and $\theta$. Agent $\theta$ exhibits a stronger precautionary saving motive than agent $\hat{\theta}$ if $\omega(\cdot,\theta)$ is a concave transformation of $\omega(\cdot,\hat{\theta})$ and $u^*_1(\cdot,\theta)$  is a convex transformation of $u^*_1(\cdot,\hat{\theta})$.
\end{co}

%Following \eqref{measure-pru}, the global measure of strength of precautionary saving motive is given by $-\frac{g_{11}(F(z),\theta)}{g_{1}(F(z),\theta)}f(z)-\frac{u^*_{111}(z,\theta)}{u^*_{11}(z,\theta)}$. 

\citet{kimball2009} study how risk aversion and intertemporal substitution affect the strength of the precautionary saving motive in a two-period model with Kreps--Porteus preferences (\citet{kp}), in which the agent's utility is given by
\begin{align}\label{kp}
\nu(c)+ \phi (\mathbb{E}_{F} u^{*}(z)),
\end{align}
where $c$ denotes first-period consumption, $F$ is the distribution of second-period consumption, $\nu$ is the first-period utility function, $u^{*}$ is the second-period Bernoulli utility function, and $\phi$ is an increasing function. The function $u^{*}$ captures the agent's risk attitudes in the second period and the function $\phi$ offers a separation between intertemporal substitution and risk aversion. In particular, a nonlinear function $\phi$ indicates departure from the standard intertemporal expected utility maximization. Assuming $\phi$ and $u^{*}$ are sufficiently smooth, the representation \eqref{kp} is a special case of our formulation where the second-period utility is nonlinear in probabilities. Our analysis above thus applies to this Kreps–Porteus representation where the local utility function for the second period is 
\begin{align}\label{local-kp}
u(z,F)=u^{*}(z) \phi_{1} (\mathbb{E}_{F} u^{*}(z)). 
\end{align}
Plugging this local utility function to the measure  $-\frac{du_{11}(y+t,F_{y+\tilde{t}})/d y}{ du_1(y+t,F_{y+\tilde{t}})/dy}$ derived from Proposition \ref{prop-precaut-frechet}, we can examine how risk  and intertemporal preferences affect the strength of the precautionary saving motive. In particular, we provide sufficient conditions under which an agent exhibits a positive precautionary saving motive.\footnote{Corollary \ref{co-kp} is consistent with \citet[Proposition 67]{gollierbook}. Within our framework, we can also recover a local measure of the strength of the precautionary saving motive and derive results analogous to Propositions 3 and 4 in \citet{kimball2009}; details are omitted.}

\begin{co}\label{co-kp}
Assume the agent has Kreps--Porteus preferences described by \eqref{kp}. The agent exhibits a positive precautionary saving motive if $\nu_{11}\leq 0$, $u^*_{11}< 0$, $u^*_{111}\geq 0$, and $\phi_{11}\leq 0$.
\end{co}

%Note that 
%\begin{align*}
%s^{*}(F_{y+\tilde{t}})\in \argmax_{s\in [0,w]} \int v(w-s, s+y+t) dF_{\tilde{t}}(t).
%\end{align*}
%If $v_{22}\leq 0$, then $s^{*}(F_{y+\tilde{t}})$ is decreasing in $y$. That is, the agent saves less when second period income is higher. A standard assumption is $-\frac{v_{22}(c_{1}, c_{2})}{v_{2}(c_{1}, c_{2})}$ is decreasing in $c_{2}$. Given these two assumptions, a sufficient condition for the induced preference to be DARA is $-\frac{v_{22}(c_{1}, c_{2})}{v_{2}(c_{1}, c_{2})}$ is decreasing in $c_{1}$. However, this condition may not be satisfied. A reasonable sufficient condition for DARA to hold is $-\frac{v_{22}(c_{1}, c_{2})}{v_{2}(c_{1}, c_{2})}$ is decreasing in $c_{2}$, $-\frac{v_{22}(w-s, s+z)}{v_{2}(w-s, s+z)}$ is increasing in $s$, and $s^{*}(F_{y+\tilde{t}})$ is decreasing in $y$. However, this condition might be too strong. 

\section{Conclusion}\label{sec-con}

Researchers have developed a plethora of MCS results that apply to problems when decision-makers are EU.  We focus on demonstrating techniques that allow us to generalize these results. We focus on conditions that correspond to Section III of \cite{athey}, as we assume that both the action and parameter spaces are scalars.  We do so because the primary applications for comparative statics under risk are covered by these formulations.  One may want to generalize our approach (e.g., allowing for multi-dimensional action or parameter sets) or specialize (e.g., assume quasiconcave returns or a Spence-Mirlees single-crossing condition), which correspond to other existing results in the literature (including other sections of \cite{athey}).  Our techniques, point the way how one can do this in a straightforward manner.

%Moreover, we do not impose additional conditions beyond log-spm (which would correspond to SC2 in the EU world) on the local utility function.  

\begin{table}[h]
\small
\begin{tabular}{|l|l|l|l|l|l|}
\hline
\textbf{Prop.} & \begin{tabular}[c]{@{}l@{}}\textbf{Assumption} \\ \textbf{on $H(s; x)$}\end{tabular}     & \begin{tabular}[c]{@{}l@{}}\textbf{Assumption} \\ \textbf{on $v$}\end{tabular}                        & \begin{tabular}[c]{@{}l@{}}\textbf{Assumption} \\ \textbf{on $V$} \end{tabular}                                      & \textbf{Conclusion}                                                       & \textbf{EU Result} \\ \hline \hline

\ref{prop-case1-w} & \begin{tabular}[c]{@{}l@{}}$H$ is SC2 in $(-x;s)$,\\ differentiable in $x$\end{tabular}   & \begin{tabular}[c]{@{}l@{}}$v_1 \geq 0$ and log-spm\\ in $(s, \theta)$ $\forall x$\end{tabular}           & \begin{tabular}[c]{@{}l@{}} Hadamard \end{tabular}  & \begin{tabular}[c]{@{}l@{}}$U_1$ is SC1\\ in $\theta$\end{tabular}  & \begin{tabular}[c]{@{}l@{}}{\cite{athey}} \\ {Prop. 2 (ii)}\end{tabular} \\ \hline

\ref{prop-body-ext1} & \begin{tabular}[c]{@{}l@{}}$h$ is SC2 in $(x;s)$,\\ differentiable in $x$ \end{tabular}   & \begin{tabular}[c]{@{}l@{}}$v\geq 0$ and log-spm\\ in $(s,\theta)$ $\forall x$ \end{tabular}           & \begin{tabular}[c]{@{}l@{}} Hadamard \end{tabular}  & \begin{tabular}[c]{@{}l@{}}$U_1$ is SC1\\ in $\theta$\end{tabular}  & \begin{tabular}[c]{@{}l@{}}{\cite{athey}} \\ {Prop. 2 (i)}\end{tabular} \\ \hline

\ref{prop-body-ext1} & \begin{tabular}[c]{@{}l@{}}$\int_{-\infty}^s H(t;x)dt$ is SC2 in $(-x; s)$,\\ $H$ is differentiable in $x$, \\ $\int s d(H(s;\hat{x})-H(s;x))=0$ \end{tabular}   & \begin{tabular}[c]{@{}l@{}}$-v_{11}\geq 0$, and log-spm\\ in $(s,\theta)$ $\forall x$.\end{tabular}           & \begin{tabular}[c]{@{}l@{}}Hadamard\end{tabular}  & \begin{tabular}[c]{@{}l@{}}$U_1$ is SC1\\ in $\theta$\end{tabular}  & \begin{tabular}[c]{@{}l@{}}{\cite{athey}} \\ {Prop. 2 (iii)}\end{tabular} \\ \hline \hline

\ref{prop-case1} & \begin{tabular}[c]{@{}l@{}}{$H$ is SC2} {in $(-x; s)$}\end{tabular} & \begin{tabular}[c]{@{}l@{}}{$v_1 \geq 0$ and log-SPM}\\{in $(s, \theta, \alpha)$  for all $\hat{x}>x$
}\end{tabular}   & Gateaux                            & \begin{tabular}[c]{@{}l@{}}{$U$ is SC2}\\ {in $(x; \theta)$}\end{tabular} & \begin{tabular}[c]{@{}l@{}}{\cite{athey}} \\ {Prop. 2 (ii)}\end{tabular}  \\ \hline

\ref{prop-case1} & \begin{tabular}[c]{@{}l@{}}{$h(s;x)$ is SC2}  {in $(x;s)$}\end{tabular} & \begin{tabular}[c]{@{}l@{}}{$v\geq 0$, and log-spm}\\ {in $(s,\theta, \alpha)$ for all $\hat{x}>x$}\end{tabular}   & Gateaux                            & \begin{tabular}[c]{@{}l@{}}{$U$ is SC2}\\ {in $(x; \theta)$}\end{tabular} & \begin{tabular}[c]{@{}l@{}}{\cite{athey}} \\ {Prop. 2 (i)}\end{tabular} \\ \hline

\ref{prop-case1} & \begin{tabular}[c]{@{}l@{}}$\int_{-\infty}^s H(t;x)dt$ is SC2  in $(-x; s)$\\ $\int s d(H(s;\hat{x})-H(s;x))=0$ \end{tabular}   & \begin{tabular}[c]{@{}l@{}}$-v_{11}\geq 0$, and log-spm\\ in $(s,\theta, \alpha)$  for all $\hat{x}>x$\end{tabular}            & Gateaux                            &  \begin{tabular}[c]{@{}l@{}}{$U$ is SC2}\\ {in $(x; \theta)$}\end{tabular} & \begin{tabular}[c]{@{}l@{}}{\cite{athey}} \\ {Prop. 2 (iii)}\end{tabular} \\ \hline \hline

\ref{prop-U-spm} & \begin{tabular}[c]{@{}l@{}}$H$ increases in $x$  w.r.t. FOSD\end{tabular}    & \begin{tabular}[c]{@{}l@{}}$v$ is SPM in $(s, \theta$)\\ for all $\alpha \in [0,1]$\end{tabular} & Gateaux                                  & \begin{tabular}[c]{@{}l@{}}$U$ is SPM\\ in $(x, \theta)$\end{tabular} & \begin{tabular}[c]{@{}l@{}}\end{tabular} \\ \hline

\end{tabular}
\normalsize
\smallskip 

\caption{Summary and Comparison of our Results}\label{tab:sum}
\end{table}

Table \ref{tab:sum} summarizes our results from Section \ref{sec-MCS-dist}.  It highlights two things.  First, rows one to three correspond in a natural way (in terms of the assumptions in the second, third, and fourth columns) to rows four to six.  To obtain MCS, the results in the first three rows require stronger conditions on $V$ in terms of differentiability and quasiconcavity than those in rows four through six. They compensate by imposing weaker conditions on $v$; specifically, they require log-spm in $(s,\theta)$ for each fixed $x$, rather than log-spm in $(s,\theta,\alpha)$ for all pairs $\hat{x}>x$.

Second, it shows that if we compare our results to the corresponding results from \cite{athey}, we identify two natural ways to extend her conditions, captured by Propositions \ref{prop-case1-w}/\ref{prop-body-ext1} on the one hand and Proposition \ref{prop-case1} on the other.  Both ways, under the assumption of EU, deliver her Proposition 2.  The conditions in Propositions in \ref{prop-case1-w}/\ref{prop-body-ext1} and \ref{prop-case1} do not nest each other, but rather they trade off assumptions about the smoothness of preferences with the strength of log-supermodularity on local utility functions.  \ref{prop-case1} is more clearly the natural generalization of \cite{athey}, Section \ref{sec-apps} shows that in fact \ref{prop-case1-w}, although perhaps a less intuitive generalization, is far more useful in applications (and easier to check).  This is because the stronger differentiability assumption in Proposition \ref{prop-case1-w} is typically met by almost all widely used non-EU preferences, and the super-modularity conditions are far easier to verify than those in Proposition \ref{prop-case1}.\footnote{\cite{athey} notes that the results which correspond to Proposition 2 (i) and (iii) are novel to her, while (ii) dates back to \cite{diamond1974increases}.}

The key result for us in terms of applications is Proposition \ref{prop-case1-w}, which is what we leverage for all three examples in Section \ref{sec-apps}. Although we focus on the case where $U$ is quasiconcave, this assumption can be relaxed, and we could instead derive the results for each application using Proposition \ref{prop-case1}. The insight that is harder to glean from the abstract results in Section \ref{sec-MCS-dist} but immediate from the propositions in Section \ref{sec-apps} is that our conditions on the local utility functions imply that, in general, for MCS we are concerned about the interaction between the ``direction'' of the local utility function (i.e., the direction of the derivative that generates the local utility) and the location at which the derivative is taken.  This is a feature that does not exist in EU (because the directional derivatives do not depend on the location --- they are the Bernoulli utility functions).  Moreover, it is a feature that is not present in the analysis of risk preferences for non-EU utilities pioneered by \cite{machina1982}.  Thus, our analysis highlights novel intuitions that exist because of the interaction of issues arising from MCS under non-EU.  

We close with a discussion of two caveats.  First, our conditions are sufficient, not necessary (recall the EU analogues in \cite{athey} are also sufficient, but not necessary).  However, the sufficient conditions we derive are tight in the sense that their violation permits the construction of counterexamples where MCS fails. Consider, for instance, the D-DARA condition. In the context of the quadratic preferences~\eqref{quadratic} analyzed in Section~\ref{sec-ln}, and under the parametrization~\eqref{quadratic-p}, D-DARA is satisfied if and only if $\alpha \geq \beta$. Conversely, when this condition is violated (i.e., $\alpha < \beta$), we can construct instances where the optimal investment decreases in wealth, thereby violating monotonicity.\footnote{More specifically, with $\alpha = 0.42$, $\beta = 0.57$, $r = 1$, and a two-point distribution where $s = 0.5$ and $s = 2$ each occur with probability $0.5$, we obtain a violation of monotonicity: when $w = 1.86$, the maximizer is $x^*(w) = 0.944$, whereas when $\hat{w} = 2.49$, the maximizer is $x^*(\hat{w}) = 0.941$.} Similarly, Corollary~\ref{prop-rdu-pos} establishes sufficient conditions for an RDU agent to exhibit a positive precautionary saving motive. \citet[Proposition~2]{chateauneuf2016precautionary} show that these conditions are also necessary.

Second, our results depend on the preferences being smooth enough.  However, in addition to EU, many non-EU preferences satisfy Hadamard differentiability: this includes Quadratic (\cite{chew1991mixture}), Weighted (\cite{hong1983generalization}, Implicit Weighted (\cite{hong1985}), as well as Rank-Dependent (\cite{quiggin1982}) and Cumulative Prospect Theory (\cite{tversky1992advances}) (so long as the weighting function has a bounded derivative on its domain), Koszegi-Rabin's Choice Acclimating Personal Equilibrium (\cite{kHoszegi2007reference}).\footnote{See \cite{chew1987} for nice discussion of many functional forms} Of course, our model also excludes some preferences; for example Disappointment Aversion (\cite{gul1991}) preferences are not even Gateaux, as are Cautious Expected Utility (\cite{cerreia2015cautious}).  Thus, although our results can speak to many non-EU models, more work remains to be done to extend them further.

\section{Appendix: Preliminary Lemmas}\label{sec-lemma}

Our MCS results rely on the following two technical lemmas. In particular, Lemma \ref{lm-tsc} provides the common mathematical tool used to address both the case in which actions affect distributions (discussed in Section \ref{sec-MCS-dist}) and the case in which actions affect the utility function (addressed in Appendix \ref{sec-MCS-u}).

\begin{lemma}\label{lm-tspm}
Let $v: S\times Y\times A\rightarrow\mathbb{R}$. Suppose $v(s,y,a)$ is supermodular in $(s,y)$ for all $a\in A$. Then 
\begin{align}\label{spm}
\int_A \int_S (v(s,\hat{y},a)-v(s,y,a)) d(F(s)-G(s))da\geq 0\quad\forall F\succsim_{FOSD} G, \forall \hat{y}>y. 
\end{align}
\end{lemma}
\begin{proof}
Since $v(s,y,a)$ is supermodular in $(s,y)$, $v(s,\hat{y},a)-v(s,y,a)$ is increasing in $s$ for all $a$. Thus, 
\begin{align*}
&\int_S  (v(s,\hat{y},a)-v(s,y,a)) d(F(s)-G(s))\geq 0\quad\forall a\in A, \forall F\succsim_{FOSD} G.
\end{align*}
This implies \eqref{spm}. 
\end{proof}

It is well known that the sum of supermodular functions is supermodular and the product of log-spm functions is log-spm. However, the sum of single crossing functions is not necessarily a single crossing function. \citet{quah2012} identify a condition, signed-ratio monotonicity, under which the single crossing property is preserved with aggregation. 

\begin{definition}
Let $f$ and $g$ be two SC1 functions defined on the partially ordered set $(S,\geq)$. We say that $f$ and $g$ satisfy \textbf{signed-ratio monotonicity} if they satisfy the following conditions: \\
(i) at any $s\in S$ such that $g(s)<0$ and $f(s)>0$, we have $-\frac{g(s)}{f(s)}\geq -\frac{g(\hat{s})}{f(\hat{s})}$ for all $\hat{s}>s$; and\\
(ii) at any $s\in S$ such that $f(s)<0$ and $g(s)>0$, we have $-\frac{f(s)}{g(s)}\geq -\frac{f(\hat{s})}{g(\hat{s})}$ for all $\hat{s}>s$.
\end{definition}

The next lemma follows from Theorem 2 of \citet{quah2012}.

%\footnote{Lemma 5 of \citet{athey} is a special case of Lemma \ref{lm-tsc}. } 

\begin{lemma}\label{lm-tsc}
Let $l, k: S\times \Theta \times  A\rightarrow\mathbb{R}$ and $K(\theta)=\int_A \int_S l(s,\theta,a)k(s,\theta,a) dsda$. Then $K(\theta)$ is SC1 if $k$ is log-spm in $(s,\theta,a)$ a.e. and $l$ satisfies the following:
\begin{itemize}
\item [(i)] $l$ is SC1 in $(s,\theta,a)$ a.e.,
\item [(ii)] the functions $l(s,\theta,a)$ and $l(\hat{s},\theta,\hat{a})$ of $\theta$ satisfy signed-ratio monotonicity for all $(\hat{s},\hat{a})>(s,a)$,
\item [(iii)] the functions $l(s,\theta,a)$ and $l(s,\theta,\hat{a})$ of $s$ satisfy signed-ratio monotonicity for all $\hat{a}>a$ and all $\theta$, 
\item [(iv)] the functions $l(s,\theta,a)$ and $l(\hat{s},\theta,a)$ of $a$ satisfy signed-ratio monotonicity for all $\hat{s}>s$ and all $\theta$. 
\end{itemize}
If $A\subseteq \mathbb{R}$, in the special case where $l$ is independent of $a$, the result of the lemma continues to hold if $k$ is log-spm in $(s,\theta,-a)$ a.e.  because conditions (i)-(iv) are trivially satisfied w.r.t. $-a$. Similarly, if $k$ is independent of $a$, the result of the lemma continues to hold  if $a$ is replaced with $-a$ in conditions (i)-(iv).
\end{lemma}

\section{Appendix: Proofs for Results in the Main Text}\label{sec-body-proofs}

\begin{proof}[Proof of Lemma \ref{lm-IDO}]
Fix $\hat{\theta}>\theta$. Take any $\hat{x}>x$  such that $U(\hat{x},\theta)\geq U(x',\theta)$ for all $x'\in [x,\hat{x}]\cap X$. We first show that  $U(\hat{x},\hat{\theta})-U(x,\hat{\theta})\geq 0$. Since $U(x,\theta)$ is quasiconcave in $x$, we have $U(x',\theta)$ weakly increasing from $x$ to $\hat{x}$. Since $U$ is differentiable in $x$, this means $U_1(x',\theta)\geq 0$ for all $x\leq x'<\hat{x}$. Since $U_1(x',\theta)$ is SC1 in $\theta$, this implies that $U_1(x',\hat{\theta})\geq 0$  for all  $x\leq x'<\hat{x}$. As a result, $U(\hat{x},\hat{\theta})-U(x,\hat{\theta})\geq 0$, as desired.

If $U(\hat{x},\theta)> U(x,\theta)$, we need to show  $U(\hat{x},\hat{\theta})-U(x,\hat{\theta})>0$. If $U(\hat{x},\theta)> U(x,\theta)$, then the set $\{x\leq x'\leq \hat{x}|U_1(x',\theta)>0\}$ has positive measure. Since  $U_1(x',\theta)$ is SC1 in $\theta$, it means that the set $\{x\leq x'\leq \hat{x}|U_1(x',\hat{\theta})>0\}$ has positive measure. As a result, $U(\hat{x},\hat{\theta})-U(x,\hat{\theta})>0$, as desired.
\end{proof}

\begin{proof}[Proofs of Propositions \ref{prop-case1-w} and \ref{prop-body-ext1}]
It follows from \eqref{eq-6.2} and integration by parts that
\begin{align*}
    U_1(x,\theta)&=\int v(s,\theta, x)h_2(s;x)ds=-\int v_1(s,\theta, x)H_2(s;x)ds\\
    &=v_1(\overline{s},\theta,x)\int sh_2(s;x)ds+\int v_{11}(s,\theta,x)\int_{-\infty}^s H_2(t;x)dt ds.
\end{align*}
Propositions  \ref{prop-case1-w} and \ref{prop-body-ext1} follow directly from the equalities above and Lemma \ref{lm-tsc}.
\end{proof}

\begin{proof}[Proof of Proposition \ref{prop-case1}]
Fix $\hat{x}>x$. If Assumption (1) holds, then $v$ is log-spm and $h(s,\hat{x})-h(s;x)$  is SC1. It follows from \eqref{U-v} and Lemma \ref{lm-tsc} that $U(\hat{x},\theta)-U(x,\theta)$ is SC1 in $\theta$. If Assumption (2) from Proposition \ref{prop-case1} holds, applying integration by parts to \eqref{U-v} yields
\begin{align*}
U(\hat{x},\theta)-U(x,\theta)&=\int_0^1\int v_1(s,\theta, \alpha)(H(s;x)-H(s;\hat{x}))dsd\alpha,
\end{align*}
Applying Assumption (2) and Lemma \ref{lm-tsc} again yields the desired result. The proof for Assumption (3) from Proposition \ref{prop-case1} is obtained by applying integration by parts to \eqref{U-v} twice and Lemma \ref{lm-tsc}.   
\end{proof}

\begin{proof}[Proof of Proposition \ref{prop-U-spm} ]
It follows from \eqref{U-v}   that for all $\hat{\theta}>\theta$,
\begin{align*}
&U(\hat{x},\hat{\theta})-U(x,\hat{\theta})-\Big(U(\hat{x},\theta)-U(x,\theta)\Big)\\
=&\int_0^1\int (v(s,\hat{\theta}, \alpha)-v(s,\theta, \alpha) )d(H(s;\hat{x})-H(s;x))d\alpha.
\end{align*}
Since $v$ is supermodular in $(s,\theta)$ for all $\alpha$ and $H$ is increasing in $x$, Lemma \ref{lm-tspm} implies that the above is nonnegative and, hence, $U$ is supermodular. 
\end{proof}

\begin{proof}[Proof of Proposition \ref{prop-asset3-frechet}]
Let $z=wr+x(s-r)$ which follows the distribution $H(z;x)=F_{\tilde{s}}(\frac{z-wr}{x}+r)$. Note that for any $\hat{x}>x$, 
\begin{align*}
-H(z;\hat{x})-(-H(z;x)) &= F_{\tilde{s}}\left(\frac{z-wr}{x}+r\right) - F_{\tilde{s}}\left(\frac{z-wr}{\hat{x}}+r\right) \notag \\
&\begin{cases}
=0, & \text{if } z\leq (w-\hat{x})r,\\
<0, & \text{if } (w-\hat{x})r<z<wr,\\
=0, & \text{if } z= wr,\\
>0, & \text{if } wr<z<\hat{x}(\overline{s}-r)+wr,\\
=0, & \text{if } z\geq \hat{x}(\overline{s}-r)+wr.
\end{cases}
\end{align*}
Thus, $H(z;x)$ is SC2 in $(-x;z)$ a.e. on $[ (w-\hat{x})r,\hat{x}(\overline{s}-r)+wr]$.  Combining this observation with the log-spm of $\overline{u}_1(z,\theta,F)$  and applying Proposition \ref{prop-case1-w} and Lemma \ref{lm-IDO} yields the desired result.
\end{proof}

\begin{proof}[Proof of Corollary \ref{co-chew1}]
By \eqref{rdu-local}, the local utility of an RDU agent with  risk parameter $\theta$ is  $\overline{u}(z,\theta,F)=\int_{\underline{z}}^z \omega_1(F(t), \theta)du^*(t,\theta)$.  Thus, $\overline{u}_1(z,\theta,F)=\omega_1(F(z),\theta)u^*_1(z,\theta)$ exists a.e., in particular,  at every continuity point of $F$. Since $F(z)$ increases in $z$ and both $\omega_1$ and $u^*_1$  are log-spm,  $\overline{u}_1(z,\theta,F)$ is log-spm in $(z,\theta)$. By Lemma 3 in \citet{chew1987}, the  conditions $\omega_{1}> 0$ and  $u^*_{11}<0$ guarantee that the objective function is differentiable and  quasiconcave in $x$.  The result then follows directly from Proposition~\ref{prop-asset3-frechet}. 
\end{proof}

\begin{proof}[Proof of Proposition \ref{prop-asset2-frechet}]
Let $F(\cdot;x,w)$ be the distribution of outcomes $wr+x(s-r)$. Since $r>0$, $F(\cdot;x,\hat{w})\succsim_{FOSD} F(\cdot;x,w)$ for all $\hat{w}>w$.   Then D-DARA implies $\frac{u_{11}(wr+t, F(\cdot;x,w))}{u_{1}(wr+t, F(\cdot;x,w))}$ increases in $w$. That is, $u_{1}(wr+t, F(\cdot;x,w))$  is log-spm in $(t,w)$ for all $x$.  By the same argument as in the proof of Proposition \ref{prop-asset3-frechet}, $H(t;x)=F_{\tilde{s}}(\frac{t}{x}+r)$ is SC2 in $(-x;t)$ a.e.  Then Proposition \ref{prop-case1-w} implies the desired result. 
\end{proof}

\subsection{Appendix for Section \ref{sec-precautionary}}\label{app-precautionary}

\begin{lemma}\label{lem-con-new}
Assume $\overline{V}$ is Hadamard differentiable. Assumption (1) of Proposition \ref{prop-precaut-frechet} implies $\frac{d \overline{V}(\theta, F_{y+\tilde{t}})}{d y}$ is strictly decreasing in $y$ for all $\tilde{t}$. 
\end{lemma}
\begin{proof}
Fix $y$, $\tilde{t}$, and $\hat{s}>s$.  Let $F_{\alpha,s,\hat{s}}=(1-\alpha)F_{y+\tilde{t}+s}+\alpha F_{y+\tilde{t}+\hat{s}}$. Hadamard differentiability implies 
\begin{align*}
&\frac{d\overline{V}(\theta,F_{y+\tilde{t}+\hat{s}})}{dy}-\frac{d\overline{V}(\theta,F_{y+\tilde{t}+s})}{dy}\\
=&\int_0^1 \int (\frac{d \overline{u}(y+t+\hat{s}, \theta,F_{\alpha,s,\hat{s}})}{dy}-\frac{d \overline{u}(y+t+s, \theta,F_{\alpha,s,\hat{s}})}{dy})dF_{\tilde{t}}(t)d\alpha< 0. 
\end{align*}
The inequality follows from Assumption (1) of Proposition \ref{prop-precaut-frechet}  and $\hat{s}>s$. 
\end{proof}

\begin{proof}[Proof of Proposition \ref{prop-precaut-frechet}]
Fix $y$. Since $\overline{V}$ is Hadamard differentiable,  we have
\begin{align*}
\frac{\partial \overline{V}(\theta, F_{y+x\tilde{s}})}{\partial x}=\int s \overline{u}_1(y+xs, \theta,F_{y+x\tilde{s}})dF_{\tilde{s}}(s). 
\end{align*}
Then 
\begin{align}\label{eq-U1}
U_1(x, \theta)=-\frac{\partial^2 \overline{V}(\theta, F_{y+x\tilde{s}})}{\partial x\partial y}=\int sf_{\tilde{s}}(s)(-\frac{d \overline{u}_1(y+x s, \theta,F_{y+x\tilde{s}})}{dy})ds. 
\end{align}
Note that 
\begin{align*}
U_1(0, \theta)=-\frac{d \overline{u}_1(y, \theta,\delta_{y})}{dy} \int sf_{\tilde{s}}(s)ds. 
\end{align*}
By Assumption (1) and  $\mathbb{E}s>0$, we have $U_1(0, \theta)>0$. Thus, the optimal $x>0$. Since $sf_{\tilde{s}}(s)$ is SC1 in $s$ and $-\frac{d \overline{u}_1(y+x s, \theta,F_{y+x\tilde{s}})}{dy}$ is log-spm in $(s,\theta)$ by Assumption (2), Lemma \ref{lm-tsc} implies that $U_1(x, \theta)$ is SC1 in $\theta$. Lemma \ref{lm-IDO} implies that  the optimal choice of $x$ increases in $\theta$. 
\end{proof}

\begin{proof}[Proof of Proposition \ref{co-precautionary}]
Consider agent $\theta$ solving the maximization problem \eqref{precautionary}.  Since $\frac{d\overline{u}_1(y+t,\theta, F_{y+\tilde{t}})}{dy}<0$, Lemma \ref{lem-con-new} implies that $\frac{d \overline{V}(\theta, F_{y+\tilde{t}})}{d y}$ is strictly decreasing in $y$ for all $\tilde{t}$. By the first order condition \eqref{foc}, the risk $\tilde{\varepsilon}$ causes additional saving if $\frac{\partial \overline{V}(\theta,F_{y+\tilde{\varepsilon}})}{\partial y}\geq \frac{\partial \overline{V}(\theta,\delta_{y})}{\partial y}$  for all $y$ and zero-mean risk $\tilde{\varepsilon}$. Let $F_{\alpha,y}=(1-\alpha)\delta_y+\alpha F_{y+\tilde{\varepsilon}}$. By Hadamard differentiability, 
\begin{align*}
\overline{V}(\theta, F_{y+\tilde{\varepsilon}})- \overline{V}(\theta,\delta_{y})=\int_0^1\Big( \int \overline{u}(y+\varepsilon, \theta, F_{\alpha,y})dF_{\tilde{\varepsilon}}(\varepsilon)-\overline{u}(y, \theta, F_{\alpha,y})\Big) d\alpha. 
\end{align*} 
Thus, 
\begin{align*}
\frac{\partial \overline{V}(\theta,F_{y+\tilde{\varepsilon}})}{\partial  y}-\frac{\partial \overline{V}(\theta,\delta_{y})}{\partial y} =\int_0^1\Big( \int \frac{d\overline{u}(y+\varepsilon,\theta, F_{\alpha,y})}{dy} dF_{\tilde{\varepsilon}}(\varepsilon)-\frac{d\overline{u}(y,\theta, F_{\alpha,y})}{dy} \Big) d\alpha\geq 0. 
\end{align*}
The last inequality follows from $\frac{d\overline{u}_{11}(y+\varepsilon,\theta, F_{\alpha,y})}{dy}\geq 0$ for all $y,\varepsilon$, and $\alpha$. Thus, agent~$\theta$'s optimal saving increases.
\end{proof}

\begin{proof}[Proof of Corollary \ref{prop-rdu-pos}]
Note that the local utility function is $\overline{u}(z, \theta, F)=\int_{\underline{z}}^z \omega_1(F(s), \theta)du^*(s,\theta)$. For any $y$, $t\in [\underline{t},\overline{t}]$ and $F_{\tilde{t}}\in \Delta [\underline{t},\overline{t}]$,  we have
\begin{align*}
\frac{d \overline{u}(y+t,\theta, F_{y+\tilde{t}})}{dy}=\int^t_{\underline{t}} \omega_1(F_{\tilde{t}}(s),\theta)u^*_{11}(y+s,\theta)ds. 
\end{align*}
Since $\omega_1>0$ and $u^*_{11}<0$, we obtain that $\frac{d\overline{u}(y+t,\theta, F_{y+\tilde{t}})}{dy}$ is strictly decreasing in $t$, that is, Assumption (1) of Proposition \ref{prop-precaut-frechet} is satisfied. Note that
 \begin{align}\label{eq-1}
&\frac{d\overline{u}_1(y+t,\theta, F_{y+\tilde{t}})}{dy}=\omega_1(F_{\tilde{t}}(t),\theta)u^*_{11}(y+t,\theta)\quad \text{a.e.}
\end{align}
Then
\begin{align}\label{eq-2}
\frac{d\overline{u}_{11}(y+t, \theta, F_{y+\tilde{t}})}{dy}= \omega_{11}(F_{\tilde{t}}(t),\theta)f_{\tilde{t}}(t)u^*_{11}(y+t,\theta) +\omega_1(F_{\tilde{t}}(t),\theta)u^*_{111}(y+t,\theta)\geq 0, 
\end{align}
where $f_{\tilde{t}}$ is the pdf of $\tilde{t}$ and the inequality follows from $\omega_{11}\leq 0$, $u^*_{11}<0$, and $u^*_{111}\geq 0$. By Proposition \ref{co-precautionary},  agent $\theta$ exhibits a positive precautionary saving motive.
\end{proof}

\begin{proof}[Proof of Corollary \ref{co-prudent}]
    We first verify the objective function is  quasiconcave in $x$.  It follows from \eqref{eq-U1} that 
\begin{align*}
U_1(x, \theta)=\int sf_{\tilde{s}}(s)\omega_1(F_{\tilde{s}}(s),\theta)(-u^*_{11}(y+xs,\theta) )ds. 
\end{align*}
It is clear that $U_1$ is  decreasing in $x$ as $u^*_{111}\geq 0$. Hence, $U$ is quasiconcave in $x$. 

The assumption that $\omega(\cdot,\theta)$ is a concave transformation of $\omega(\cdot, \hat{\theta})$ and $u^*_1(\cdot,\theta)$  is a convex transformation of $u^*_{1}(\cdot, \hat{\theta})$ implies  
\begin{align*}
-\frac{\omega_{11}(z,\theta)}{\omega_1(z,\theta)}\geq -\frac{\omega_{11}(z,\hat{\theta})}{\omega_1(z,\hat{\theta})} \quad\text{and} \quad -\frac{u^*_{111}(z,\theta)}{u^*_{11}(z,\theta)}\geq -\frac{u^*_{111}(z,\hat{\theta})}{u^*_{11}(z,\hat{\theta})}\quad\forall z.
\end{align*}
Using \eqref{eq-1} and \eqref{eq-2}, we obtain
\begin{align*}
-\frac{d \overline{u}_{11}(y+t,\theta,F_{y+\tilde{t}})/d y}{ d\overline{u}_1(y+t,\theta,F_{y+\tilde{t}})/dy}\geq -\frac{d \overline{u}_{11}(y+t,\hat{\theta},F_{y+\tilde{t}})/d y}{ d\overline{u}_1(y+t,\hat{\theta},F_{y+\tilde{t}})/dy}\quad \forall y,t,\tilde{t}. 
\end{align*}
By Proposition \ref{prop-precaut-frechet}, agent $\theta$ exhibits a stronger precautionary saving motive.    
\end{proof}

\begin{proof}[Proof of Corollary \ref{co-kp}]
Since $\nu_{11}\leq 0$,  Proposition \ref{co-precautionary} proves that the risk represented by $\tilde{\varepsilon}$ will cause additional saving if $\frac{du_{11}(y+t, F_{y+\tilde{t}})}{dy} \geq 0$. By \eqref{local-kp}, we know that
\begin{align*}
&\frac{du_{11}(y+t, F_{y+\tilde{t}})}{dy} \\
=& u^*_{111}(y+t) \phi_{1} \left(\mathbb{E} \left[u^*(y+\tilde{t})\right]\right) + u^*_{11}(y+t) \phi_{11} \left(\mathbb{E} \left[u^*(y+\tilde{t})\right]\right) \mathbb{E} \left[u^*_1(y+\tilde{t})\right].
\end{align*}
Obviously, $u^*_{11}< 0$, $u^*_{111}\geq 0$, and $\phi_{11}\leq 0$ together imply that $\frac{du_{11}(y+t, F_{y+\tilde{t}})}{dy} \geq 0$, as desired. 
\end{proof}

\newpage 

\small\bibliographystyle{abbrvnat}
\bibliography{dt}

\newpage

\appendix

% If you want subsections labeled A1, A2, ... (not A.1, A.2)
\renewcommand{\thesubsection}{\thesection\arabic{subsection}}

% Theorem-like environments, numbered A1, A2, ... within each Appendix section
%\newtheorem{prop}{Proposition}[section]
\renewcommand{\theprop}{\Alph{section}\arabic{prop}}

\renewcommand{\thethm}{\Alph{section}\arabic{thm}}

\renewcommand{\theco}{\Alph{section}\arabic{co}}

\renewcommand{\thelemma}{\Alph{section}\arabic{lemma}}

% Equations numbered (A1), (A2), ... within each Appendix section
\renewcommand{\theequation}{\Alph{section}\arabic{equation}}
\setcounter{equation}{0}

\setlength{\abovedisplayskip}{5pt}
\setlength{\belowdisplayskip}{5pt}
\setlength{\abovedisplayshortskip}{5pt}
\setlength{\belowdisplayshortskip}{5pt}

\section{Actions Affect Utility}\label{sec-MCS-u}

Recall that the main text analyzes the case in which the action affects the distribution over states, while the parameter enters the utility function. In this appendix, we consider the complementary case in which these roles are reversed: the parameter affects the distribution, and the action influences utility.

Before turning to the details of this alternative setting, we first discuss a formulation that encompasses both the analysis in this appendix and that in the main text. A general formulation is given by $U(x,\theta)={V}(\theta, F(\cdot; x,\theta))$, where  \(\theta\) may enter the utility function directly and both \(x\) and \(\theta\)  may also influence the lottery over final outcomes \(F\).\footnote{Note that this formulation excludes the case where $x$ directly affects the utility function. In that situation, the difference  $U(\hat{x},\theta)-U(x,\theta)$ cannot be expressed in terms of  local utility functions. This case is studied in this Appendix just a bit later.} Recall that \(V\) denotes a preference functional over lotteries, with a given lottery induced by the pair \((x,\theta)\). For any $\hat{x}>x$,  Gateaux differentiability of ${V}$ implies 
\begin{align}\label{eq-general}
U(\hat{x},\theta)-U(x,\theta)=\int_{0}^{1} \int {u}(z,\theta, F_{\alpha,x,\hat{x}}(\cdot;\theta))d ( F(z;\hat{x},\theta)- F(z; x,\theta))d\alpha,
\end{align}
where  $F_{\alpha,x,\hat{x}}(\cdot;\theta)=(1-\alpha)F(\cdot; x,\theta)+\alpha F(\cdot;\hat{x},\theta)$ for each $\alpha\in [0,1]$ and ${u}$ is the local utility function of $V$.  By way of analogy, observe that the equivalent formulation under EU would be 
\begin{align*}
U(\hat{x},\theta)-U(x,\theta)= \int {u}^*(z,\theta)d ( F(z;\hat{x},\theta)- F(z; x,\theta)),
\end{align*}
where the Bernoulli utility does not depend on $F$, and thus we do not need to integrate over $\alpha$. However, the functional form in \eqref{eq-general} is very general, so the sufficient conditions on local utility function ${u}$ and distribution function $F$ for MCS to hold are difficult to interpret, which is why we consider the distinct cases in the main text and this Appendix.

Returning to the situation where $x$ impacts the utility and $\theta$ impacts the distribution over states, let $s$ denote a state and the distribution of states is given by pdf $g(s;\theta)$ and cdf $G(s;\theta)$. Just as in the main text of the paper, there are two cases to consider, which capture the different ways in which $x$ can impact utility.  In the first case, actions  directly affect the  local utility without affecting outcomes.  This case corresponds to the case studied in \citet{machina1989}, and we turn to replicating his results here using our framework. Slightly abuse of notation, let the outcome be $s$. Let $\tilde{V}(x, G(\cdot;\theta))$ be the preference functional with the local utility function $\tilde{u}(s,x,G)$.  The key observation is $x$ does not affect outcomes and, hence, the lottery $G(\cdot;\theta)$ is independent of $x$. The objective function in this case is $U(x,\theta)=\tilde{V}(x, G(\cdot;\theta))$.

Since we study general non-EU preferences, we do not impose a specific functional form on preferences. When the action $x$ does not affect outcomes, the difference \(U(\hat{x},\theta)-U(x,\theta)\) cannot be expressed in terms of local utility functions. Consequently, we cannot derive meaningful sufficient conditions on local utilities for this difference to be SC1 in \(\theta\). In this case, however, we can instead establish conditions under which \(U(x,\theta)\) is supermodular. To do so, rather than computing \(U(\hat{x},\theta)-U(x,\theta)\), we compute $U(x,\hat{\theta}) - U(x,\theta)$ for \(\hat{\theta} > \theta\), which can be expressed as a function of local utility functions.\footnote{We can provide conditions under which $U(x, \theta)$ is supermodular because supermodularity is symmetric in $(x,\theta)$ in the sense that 
\begin{align*}
U(\hat{x},\hat{\theta})-U(x,\hat{\theta})\geq U(\hat{x},\theta)-U(x,\theta)~\Leftrightarrow~ U(\hat{x},\hat{\theta})-U(\hat{x},\theta)\geq U(x,\hat{\theta})-U(x,\theta)
\end{align*}
and we can derive conditions for supermodularity by computing $U(x,\hat{\theta})-U(x,\theta)$ instead. However, SC2 is not symmetric in $(x,\theta)$ and, hence, the same technique is not applicable.} As a result, we can find sufficient conditions on local utilities and distribution functions for $U$ to be supermodular. This case is analyzed in Section \ref{sec-machina-spm}.

The second case is where actions affect utility through the outcomes. Let $\kappa(s,x)$ be the outcome when state $s$ is realized and action $x$ is chosen. For each choice of $x$ and each parameter $\theta$, let $F(\cdot; x, \theta)$ denote the distribution over final outcomes. The objective function in this case is $U(x,\theta)=V(F(\cdot; x,\theta))$ with the local utility function $u(\kappa(s,x),F(\cdot; x, \theta))$.

\subsection{Sufficiency for $U_1(x,\theta)$ to be SC1}

First, we consider the case in which \(U_1\) is SC1. Our analysis parallels that in Section~\ref{sec-sc1}.  As discussed above, we restrict our attention to the case where the action $x$ affects the outcomes $\kappa(s,x)$. 

As in the main text, we begin with a relatively strong differentiability assumption on $V$ --- Hadamard differentiability.  By Hadamard differentiability, 
\begin{align*}
U_1(x,\theta)=&\int  u_1(\kappa(s,x), F(\cdot;x,\theta))\kappa_2(s,x) g(s;\theta)ds,
\end{align*} 
where $\kappa_2(s,x)=\frac{\partial \kappa(s,x)}{\partial x}$.

\begin{prop}\label{prop-sc_1}
Assume \begin{enumerate}
    \item $V$ is Hadamard differentiable;
        \item $g(s;\theta)$ is log-spm a.e.;
    \item  $u(z,F)$ is strictly increasing in $z$ for all $F$, $\kappa(s,x)$ is  increasing in $s$, and $\kappa(s,x)$ has SC2 in $(x;s)$ a.e., $\kappa(s,x)$ is differentiable in $x$;
    \item $u_1(z, F)$ is log-spm in $(z,F)$ a.e. 
\end{enumerate}
Then $U_1(x,\theta)$ is SC1 in $\theta$.
\end{prop}

The assumptions and the result are parallel to Proposition~\ref{prop-case1-w}. Assumption (2) means that $\theta$ makes  higher states relatively more likely. In Assumption (3), the assumptions that $u(z,F)$ is strictly increasing in $z$ and $\kappa(s,x)$ is  increasing in $s$ are fairly innocuous. Assumption (4)  is a kind of  decreasing absolute risk aversion assumption, which says that local risk aversion is falling in the second argument of the local utilities. That is, $-\frac{u_{11}(z,F)}{u_1(z,F)}$ decreases in  $F$.   

\subsection{Sufficiency for $U$ to be SC2}\label{sec-sc2-x}
We now turn to discussing conditions that guarantee that $U$ is SC2.  Again, we restrict our attention to the case where the action $x$ affects the outcomes $\kappa(s,x)$.

%In the case where actions do not affect the outcomes, if the preference functional $\overline{V}(x, F(\cdot;\theta))$ is not explicitly known, we cannot express $\overline{V}(\hat{x}, F(\cdot;\theta))-\overline{V}(x, F(\cdot;\theta))$ in terms of local utility functions. As a result, we are not able to provide conditions on the local utility functions so that  $\overline{V}(\hat{x}, F(\cdot;\theta))-\overline{V}(x, F(\cdot;\theta))$ is SC1 in $\theta$.\footnote{We can provide conditions under which $\overline{V}(x, F(\cdot;\theta))$ is supermodular because supermodularity is symmetric in $(x,\theta)$ in the sense that 
%\begin{align*}
%U(\hat{x},\hat{\theta})-U(x,\hat{\theta})\geq U(\hat{x},\theta)-U(x,\theta)~\Leftrightarrow~ U(\hat{x},\hat{\theta})-U(\hat{x},\theta)\geq U(x,\hat{\theta})-U(x,\theta)
%\end{align*}
%and we can derive conditions for supermodularity by computing $U(x,\hat{\theta})-U(x,\theta)$ instead. However, SC2 is not symmetric in $(x,\theta)$ and, hence, the same technique is not applicable.} Therefore, in this section, we restrict our attention to the case where the action $x$ affects the outcomes $\kappa(s,x)$ and $U(x,\theta)=V(P(x,\theta))$ with local utility function $u(z,P)$. 

%For each possible outcome $z$, its density is $(1-\alpha)g(s(z);\theta)+\alpha g(\hat{s}(z);\theta)$ where $\kappa(s(z),x)=z$ and $\kappa(\hat{s}(z),\hat{x})=z$.
For any $\hat{x}>x$ and $\alpha\in [0,1]$, define $F_{\alpha,x,\hat{x}}(\theta)=(1-\alpha)F(\cdot; x,\theta)+\alpha F(\cdot; \hat{x},\theta)$.  Using the Gateaux differentiability of $V$, we obtain
\begin{align}\label{U-Frechet}
\begin{split}
U(\hat{x},\theta)-U(x,\theta)=&\int_0^1 \int ( u(\kappa(s,\hat{x}), F_{\alpha,x,\hat{x}}(\theta))-u(\kappa(s,x),  F_{\alpha,x,\hat{x}}(\theta)) ) dG(s;\theta) d\alpha. 
\end{split}
\end{align}
Analogous to the arguments in Section \ref{sec-sc2}, we need to use our extension of \citet{quah2012} to aggregate local utilities. 
 
Our first result is analogous to condition (1) of Proposition \ref{prop-case1}.

\begin{prop}\label{lm-sufficient1}
Assume 
\begin{enumerate}
    \item $V$ is Gateaux differentiable;
    \item $g(s;\theta)$ is log-spm a.e.;
    \item $u(\kappa (s,x'), F_{\alpha,x,\hat{x}}(\theta))$ is log-spm in $(s,x',\alpha,\theta)$ a.e. for  all $\hat{x}>x$. 
\end{enumerate}   
Then  $U(x,\theta)$ is SC2 in $(x;\theta)$. 
\end{prop}

The interpretation of these conditions is analogous to what we provided in the main text of the paper, modulo the switch of $x$ and $\theta$.

We can also provide conditions that do not require that the local $u$ is log-spm.  

\begin{prop}\label{prop-sc}
Assume 
\begin{enumerate}
    \item $V$ is Gateaux differentiable
 \item $g(s;\theta)$ is log-spm a.e.; 
        \item  $u(z,F)$ is strictly increasing in $z$ for all $F$, $\kappa(s,x)$ is  increasing in $s$, and $\kappa(s,x)$ has SC2 in $(x;s)$ a.e.;
    \item One of the following holds:
\begin{itemize}
\item [a.] $u(\kappa(s,x),F)$ is supermodular in $(x,F)$ a.e. for all $s$; 
\item [b.] For all $\hat{x}>x$, $u_1(z, F_{\alpha,x,\hat{x}}(\theta))$ is log-spm in $(z, \alpha,\theta)$ a.e.
\end{itemize}
\end{enumerate}
Then $U(x,\theta)$ is SC2 in $(x;\theta)$.\footnote{It follows from Lemma \ref{lm-tsc} that Proposition \ref{prop-sc} continues to hold if $\alpha$ is replaced by $-\alpha$ in Assumption (4b). Indeed,  we only need $-\frac{u_{11}(z, F_{\alpha,x,\hat{x}}(\theta))}{u_{1}(z, F_{\alpha,x,\hat{x}}(\theta))}$ to be monotone in $\alpha$. 
} 
\end{prop}

Assumptions (1) and (2) are entirely standard. Assumption (3) ensures that $u(\kappa(s,x),F)$ has SC2 in $(x;s)$ a.e. for all $F$. Assumptions (2) and (3) together imply $F_{\alpha,x,\hat{x}}(\hat{\theta}) \succsim_{FOSD} F_{\alpha,x,\hat{x}}(\theta) $ for all $ \hat{\theta}>\theta$, $\hat{x}>x$, and  $\alpha\in [0,1]$. 

Assumption (4) ensures single-crossing properties can be aggregated in our setting. Obviously, Assumption (4) is satisfied trivially under EU. Assumption  (4a) is very restrictive.  To see this, suppose there exist $\hat{x}>x$ and $s, \hat{s}$ such that $\kappa(s,\hat{x})-\kappa(s,x)<0<\kappa(\hat{s},\hat{x})-\kappa(\hat{s},x)$,  then Assumption (4a) requires for all $\hat{F}\succsim_{FOSD} F$, 
\begin{align*}
&u(\kappa(s,\hat{x}),\hat{F})-u(\kappa(s,x),\hat{F})\geq u(\kappa(s,\hat{x}),F)-u(\kappa(s,x),F)\quad \text{and}\\
&u(\kappa(\hat{s},\hat{x}),\hat{F})-u(\kappa(\hat{s},x),\hat{F})\geq u(\kappa(\hat{s},\hat{x}),F)-u(\kappa(\hat{s},x),F), 
\end{align*}
which says $u(z,F)$ is supermodular both in $(z,F)$ and $(-z,F)$, that is, $u(\hat{z},F)-u(z,F)$ is constant in $F$. Under differentiability,  an equivalent statement of Assumption (4b) is that  $-\frac{u_{11}(z, F_{\alpha,x,\hat{x}}(\theta))}{u_{1}(z, F_{\alpha,x,\hat{x}}(\theta))}$ is decreasing in $\alpha$ and $\theta$.

\subsection{Sufficiency for $U$ to be Supermodular}\label{sec-machina-spm}

We conclude this section by considering the last sufficient condition for MCS --- that $U$ is supermodular.  

As discussed previously, we cannot express $U(\hat{x},\theta)-U(x,\theta)$ in terms of local utility function if $x$ does not affect the distribution of final outcomes. However, we  can  provide sufficient conditions for $U$ to be supermodular by computing $U(x,\hat{\theta})-U(x,\theta)$.  When the action $x$ does not affect outcomes, for all $\hat{\theta}>\theta$ and  $\alpha\in [0,1]$, let $G_{\alpha,\theta,\hat{\theta}}(\cdot)=(1-\alpha)G(\cdot;\theta)+\alpha G(\cdot;\hat{\theta})$. If $\tilde{V}$ is Gateaux differentiable, we have
\begin{align}\label{eq-nooutcome}
\begin{split}
U(x,\hat{\theta})-U(x,\theta)&=\tilde{V}(x, G(\cdot;\hat{\theta}))-\tilde{V}(x, G(\cdot;\theta))\\
&=\int_0^1 \int \tilde{u}(s,x, G_{\alpha,\theta,\hat{\theta}}(\cdot))d(G(s;\hat{\theta})-G(s;\theta))d\alpha. 
\end{split}
\end{align}
When the action $x$ affects the outcomes, let $P_{\alpha, \theta,\hat{\theta}}(\cdot; x)=(1-\alpha)F(\cdot; x,\theta)+\alpha F(\cdot;x,\hat{\theta})$. That is, $P_{\alpha, \theta,\hat{\theta}}(\cdot; x)$ is the lottery that in each state $s$, the outcome is $\kappa(s,x)$ and its pdf is given by  $(1-\alpha) g(s;\theta)+\alpha g(s;\hat{\theta})$.  Then by Gateaux differentiability, we have
\begin{align}\label{eq-outcome-x}
\begin{split}
U(x,\hat{\theta})-U(x,\theta)&=V(F(\cdot;x,\hat{\theta}))-V(F(\cdot; x,\theta))\\
&=\int_0^1 \int u(\kappa(s,x), P_{\alpha, \theta,\hat{\theta}}(\cdot; x))d(G(s;\hat{\theta})-G(s;\theta))d\alpha. 
\end{split}
\end{align}
Observe from \eqref{eq-nooutcome} and \eqref{eq-outcome-x}  that in both cases, we only need to show 
\begin{align}\label{eq-bothcase}
U(x,\hat{\theta})-U(x,\theta)=\int_0^1 \int v(s,x, \alpha)d(G(s;\hat{\theta})-G(s;\theta))d\alpha
\end{align}
is increasing in $x$, where $v(s,x, \alpha)= \tilde{u}(s,x, G_{\alpha,\theta,\hat{\theta}}(\cdot))$ in the former case and $v(s,x, \alpha)=u(\kappa(s,x), P_{\alpha, \theta,\hat{\theta}}(\cdot; x))$ in the latter. Note that the key difference between the two cases is whether $x$ enters the lottery argument of the local utility function.

Thus, we look to provide a set of sufficient conditions so that $U(x,\hat{\theta})-U(x,\theta)$ increases in $x$. Because the sum of supermodular functions is supermodular, we do not need to worry about aggregation. The following result is an immediate implication of  Lemma \ref{lm-tspm}.

\begin{prop}\label{prop-machina}
Assume 
\begin{enumerate}
    \item preference functionals are Gateaux differentiable;
        \item $G(\cdot; \hat{\theta})\succsim_{FOSD} G(\cdot; \theta)$ for all $\hat{\theta}>\theta$;
    \item  $v(s, x, \alpha)$ is supermodular in $(s,x)$  for all $\hat{\theta}>\theta$ and $\alpha\in [0,1]$.
\end{enumerate}
Then $U(x,\theta)$ is supermodular in $(x,\theta)$. 
\end{prop}

The interpretation of these assumptions mirrors that of Proposition~\ref{prop-U-spm}, with the roles of $x$ and $\theta$ interchanged.  We note that (3) requires $v(s, x, \alpha)$ be supermodular in $(s,x)$ for all $\alpha\in [0,1]$. When $x$ affects the outcomes, this condition requires the primitive local utility function $u(\kappa(s,\hat{x}), F(\cdot; \hat{x},\theta))-u(\kappa(s,x), F(\cdot; x,\theta)) $ be increasing in $s$ for all $\theta$. Whether this assumption is satisfied or not depends on  how $u(z,F)$ depends on $F$ and how $\kappa(s,x)$ varies with $x$. A sufficient condition for Assumption (3)  to hold is  $u(\kappa(s,x), F)$ is supermodular in $(s,x)$ for all $F$ and $u(z,F(\cdot; x,\theta))$ is supermodular in $(z, x)$ for all $\theta$.\footnote{It is well-known that supermodularity is not preserved under monotone transformations, so supermodularity of $\kappa$ does not ensure supermodularity of $u(\kappa(s,x), F)$. }

When the action $x$ does not affect the outcomes, Proposition \ref{prop-machina} replicates Theorem (i) in \citet{machina1989}. The objective function $\tilde{V}(x, G(\cdot;\theta))$ satisfies what Machina termed ``functional separation of the probability distribution  from the control variable''.

%We now provide an example of a utility function that satisfies Assumption \ref{as-spm-x} (ii) in order to demonstrate its relevancy (details of this example, as well as all others, are provided in Appendix \ref{app-examples}).

%\begin{example}\label{ex-1}
%\textit{Suppose $x$ affects the outcomes $\kappa(s,x)$. In Appendix \ref{app-examples}, we provide sufficient conditions under which the quadratic preference functional  \eqref{quadratic} satisfies Assumption \ref{as-spm-x} (ii).}

%\end{example}

\subsection{Application: Portfolio Choice and Changes to Returns}\label{sec-app1}

Consider the same portfolio choice problem as in Section~\ref{sec-dara}. In contrast to the main text, assume now that the distribution of the random return $s$ depends on a parameter $\theta$, with cdf $G(s;\theta)$ and pdf $g(s;\theta)$. Suppose the expected return of $s$ under all possible $\theta$ exceeds the risk-free return $r$. The outcome function is then $\kappa(s,x)=wr+x(s-r)$, which satisfies SC2 in $(x;s)$. This setting corresponds to the application in \citet[p.~205]{athey} and is one of the classic problems in choice under risk. 

An immediate application of Proposition \ref{prop-sc_1} yields the next result.

\begin{prop}\label{prop-asset1-frechet}
Suppose $V$ is Hadamard differentiable and $U$ is  quasiconcave in $x$. Suppose further that $u(z,F)$ is strictly increasing in $z$ for all $F$, $u_1(z, F)$ is log-spm in $(z,F)$ a.e., and $g(s;\theta)$ is log-spm a.e. Then an increase in $\theta$ leads to an increase in the optimal choice $x$.
\end{prop}

The log-spm of $g(s;\theta)$ implies that $s$ and $\theta$ are affiliated.\footnote{\citet{milgrom1982} show that a vector of random variables is affiliated if and only if their joint density is log-spm. } For example, $\theta$ can be a signal received by the agent that is affiliated with the return $s$, so that  a higher signal  makes higher returns relatively more likely. The log-spm of $u_1(z,F)$ is equivalent to $-\frac{u_{11}(z,F)}{u_1(z,F)}$ decreasing in $F$, representing a form of decreasing absolute risk aversion---meaning the agent becomes less risk averse when evaluated at a better distribution. Proposition \ref{prop-asset1-frechet} therefore states that when the agent receives a more favorable signal about the return to the risky asset and is less risk averse when evaluated at a better distribution, then she would invest more in the risky asset.

\citet[Proposition~3]{athey} also considers a case where the Bernoulli utility depends on the parameter. Under non-EU preferences, the parameter $\theta$ naturally enters the local utility function through the distribution function. Hence, the restriction on how the utility depends on $\theta$ becomes a restriction on how the local utility depends on the distribution in our framework. The analogue of \citet{athey}'s condition~(B) in Proposition~3 is that $u_1(z,F)$ is log-spm in $(z,F)$.

\subsection{Proofs}\label{sec-MCS-u-proofs}

\begin{proof}[Proof of Proposition \ref{prop-sc_1}]
Since $g(s;\theta)$ is log-spm,  we have $G(\cdot;\hat{\theta})\succsim_{FOSD} G(\cdot;\theta)$ and, hence, $F(\cdot; x,\hat{\theta})\succsim_{FOSD} F(\cdot;x,\theta)$ for all $\hat{\theta}>\theta$.  Combining this with the log-spm of $u_1$, we know that $u_1(z, F(\cdot; x,\theta))$ is log-spm in $(z,\theta)$. Since  $u(z,F)$ is strictly increasing in $z$ for all $F$ and $\kappa(s,x)$ is  increasing in $s$, we also obtain that $u_1(\kappa(s,x), F(\cdot; x,\theta))$ is log-spm in $(s,\theta)$. The SC2 of $\kappa(s,x)$  implies $\kappa_2(s,x)$ is SC1 in $s$. Then the result is an immediate implication of Lemma \ref{lm-tsc}. 
\end{proof}

\begin{proof}[Proof of Proposition \ref{lm-sufficient1}]
By Assumptions (2) and (3), the function $u(\kappa (s,x'), F_{\alpha,x,\hat{x}}(\theta))g(s;\theta)$ is log-spm a.e. By Lemma 2 in \citet{athey}, the function $A(x',\theta)= \int_0^1 \int u(\kappa (s,x'), F_{\alpha,x,\hat{x}}(\theta))g(s;\theta)ds d\alpha$ is log-spm in $(x',\theta)$. This implies 
\begin{align}\label{ineq-H}
\frac{A(\hat{x},\hat{\theta})}{A(x,\hat{\theta})}\geq \frac{A(\hat{x},\theta)}{A(x,\theta)}\quad\forall \hat{\theta}>\theta.
\end{align}
By \eqref{U-Frechet}, we have $U(\hat{x},\theta)-U(x,\theta)= A(\hat{x},\theta)-A(x,\theta).$ Then for any $\hat{\theta}>\theta$, if $U(\hat{x},\theta)-U(x,\theta)= A(\hat{x},\theta)-A(x,\theta)\geq (>) 0$, it follows from \eqref{ineq-H} that $U(\hat{x},\hat{\theta})-U(x,\hat{\theta})= A(\hat{x},\hat{\theta})-A(x,\hat{\theta})\geq (>) 0$. 
\end{proof}

\begin{lemma}\label{lm-s0}
Suppose Assumption (3) in Proposition \ref{prop-sc} holds. Fix $\hat{x}>x$. Then for all  $F$, $m(s,F)=u(\kappa(s,\hat{x}), F)-u(\kappa(s,x), F)$ is  SC1 in $s$. Moreover, if  there exist $s^*>s_*$ such that  
\begin{align}\label{ineq-s*}
\kappa(s^*,\hat{x})-\kappa(s^*,x)>0> \kappa(s_*,\hat{x})-\kappa(s_*,x),
\end{align}
then there exists $s_0\in [s_*,s^*]$ such that $m(\hat{s},F)\geq 0\geq m(s,F)$ for all $F$ and $\hat{s}>s_0>s$. 
\end{lemma}
\begin{proof}
Since $\kappa$ is SC2 in $(x;s)$ and $u$ is strictly increasing in $z$, the local utility function $u(\kappa(s,x),F)$ is SC2 in $(x;s)$ for all $F$. Thus, $m(s, F)$ is SC1 in $s$. 

Suppose such $s^*,s_*$ exist. Since $\kappa$ is SC2 in $(x;s)$, there exists $s_0\in [s_*,s^*]$ such that  
\begin{align*}
\kappa(\hat{s},\hat{x})-\kappa(\hat{s},x)\geq 0\geq  \kappa(s,\hat{x})-\kappa(s,x)\quad\forall \hat{s}>s_0>s. 
\end{align*}
Since $u$ is strictly increasing, we have $m(\hat{s},F)\geq 0\geq m(s,F)$ for all $F$ and $\hat{s}>s_0>s$. 
\end{proof}

\begin{proof}[Proof of Proposition \ref{prop-sc}]
Fix $\hat{x}>x$. If there do not exist $s^*>s_*$ such that \eqref{ineq-s*} holds, which means  either $m(s,F)\geq 0$ for all $s$ and $F$, or  $m(s,F)\leq 0$ for all $s$ and $F$, then the result holds trivially. We thus assume such $s^*>s_*$ exist. 

Suppose first Assumption (4a) holds. For any $\hat{\theta}>\theta$, the fact that $g(s;\theta)$ is log-spm a.e. implies  $F_{\alpha,x,\hat{x}}(\hat{\theta})\succsim_{FOSD} F_{\alpha,x,\hat{x}}(\theta)$ for all $\alpha$.  By Assumption (4a), we know that $m(s,F_{\alpha, x,\hat{x}}(\theta))$ is an increasing function of $\theta$. By Lemma \ref{lm-s0},  $m(s,F) $ is SC1 in $s$ for all $F$ and there exists $s_0$ such that  $m(\hat{s},F)\geq 0\geq m(s,F)$ for all $F$ and $\hat{s}>s_0>s$. Take $s<s_0<\hat{s}$. Since $g(s;\theta)$ is log-spm a.e.,
\begin{align*}
\frac{g(\hat{s};\hat{\theta})}{g(s;\hat{\theta})}\geq \frac{g(\hat{s};\theta)}{g(s;\theta)}\quad\forall\hat{\theta}>\theta. 
\end{align*}
Combining this with the fact that $m(s,F_{\alpha, x,\hat{x}}(\theta))$ is increasing in $\theta$, we obtain
\begin{align*}
-\frac{m(s,F_{\alpha, x,\hat{x}}(\theta))g(s;\theta)}{m(\hat{s}, F_{\hat{\alpha},x,\hat{x}}(\theta))g(\hat{s};\theta)}\geq -\frac{m(s, F_{\alpha,x,\hat{x}}(\hat{\theta}))g(s;\hat{\theta})}{m(\hat{s}, F_{\hat{\alpha},x,\hat{x}}(\hat{\theta}))g(\hat{s};\hat{\theta})} \quad\forall \hat{\alpha}, \alpha, \forall\hat{\theta}>\theta. 
\end{align*}
Thus, functions  $m(s,F_{\alpha, x,\hat{x}}(\theta))g(s;\theta)$ and $m(\hat{s}, F_{\hat{\alpha},x,\hat{x}}(\theta))g(\hat{s};\theta)$ of $\theta$ satisfy   the signed-ratio monotonicity. Then it follows from \eqref{U-Frechet} and Theorem 1 in \citet{quah2012} that $U(\hat{x},\theta)-U(x,\theta)$ is SC1 in $\theta$.

Suppose  instead Assumption (4b) holds. It follows from \eqref{U-Frechet} that we only need to prove $m(s,F_{\alpha,x,\hat{x}}(\theta))$ satisfies conditions (i)-(iv) in  Lemma \ref{lm-tsc}. Since $u$ is strictly increasing in $z$, whether $m$ is positive, negative, or zero is determined completely by $s$. By Lemma \ref{lm-s0},  $m(s,F)$ is SC1 in $s$. Thus, $m(s,F_{\alpha, x,\hat{x}}(\theta))$  has SC in $(s, \alpha, \theta)$, i.e., condition (i) in  Lemma \ref{lm-tsc} is satisfied. Also, since $m(s,F_{\alpha, x,\hat{x}}(\theta))$ and $m(s, F_{\hat{\alpha},x,\hat{x}}(\theta))$ must have the same sign for any $\hat{\alpha}>\alpha$ and $\theta$, functions $m(s,F_{\alpha, x,\hat{x}}(\theta))$ and $m(s, F_{\hat{\alpha},x,\hat{x}}(\theta))$ of $s$ satisfy the signed-ratio monotonicity trivially. Thus, condition (iii) in  Lemma \ref{lm-tsc} is satisfied. 

We now verify Lemma \ref{lm-tsc} (ii) for all $(\hat{s},\hat{\alpha})>(s,\alpha)$, functions $m(s,F_{\alpha, x,\hat{x}}(\theta))$ and $m(\hat{s}, F_{\hat{\alpha},x,\hat{x}}(\theta))$ of $\theta$ satisfy the signed-ratio monotonicity and (iv) for all $\hat{s}>s$, functions $m(s,F_{\alpha, x,\hat{x}}(\theta))$ and $m(\hat{s}, F_{\alpha,x,\hat{x}}(\theta))$ of $\alpha$ satisfy the signed-ratio monotonicity. Define $h(a,b, F)=\int_a^b u_1(y, F)dy$. Since $u_1(y, F_{\alpha,x,\hat{x}}(\theta))$ is log-spm in $(y, \alpha,\theta)$, $h(a,b, F_{\alpha,x,\hat{x}}(\theta))$ is log-spm in  $(a,b, \alpha,\theta)$ for all $a<b$. For any $s\geq s_0$, we know that $\kappa(s,\hat{x})\geq \kappa(s,x)$. Since $\kappa$ increases in $s$, this implies $m(s,F_{\alpha,x,\hat{x}}(\theta))$ is log-spm in $(s, \alpha, \theta)$ for all $s\geq s_0$, where $m(s,F_{\alpha,x,\hat{x}}(\theta))=h(\kappa(s,x),\kappa(s,\hat{x}), F_{\alpha,x,\hat{x}}(\theta))$.  If $s<s_0$, then $m(s,F_{\alpha,x,\hat{x}}(\theta))=-h(\kappa(s,\hat{x}),\kappa(s,x), F_{\alpha,x,\hat{x}}(\theta))$ and $-m(s,F_{\alpha,x,\hat{x}}(\theta))$ is log-spm in $(s,  \alpha,\theta)$. 

Let $s_0$ be defined as in Lemma \ref{lm-s0}. Take $s<s_0<\hat{s}$, $\theta<\hat{\theta}$, $\alpha<\hat{\alpha}$. We know that $m(s,F_{\alpha, x,\hat{x}}(\theta))\leq 0$ and $m(\hat{s}, F_{\hat{\alpha},x,\hat{x}}(\theta))\geq 0$. Since  $m(s',F_{\alpha,x,\hat{x}}(\theta))$ is log-spm in $(s',\alpha, \theta)$ for all $s'\geq s_0$, we obtain
\begin{align}\label{ineq-ratio1}
\frac{m(\hat{s},F_{\hat{\alpha},x,\hat{x}}(\hat{\theta}))}{m(\hat{s},F_{\hat{\alpha},x,\hat{x}}(\theta))}\geq \limsup _{s'\downarrow s_0}\frac{m(s',F_{\alpha,x,\hat{x}}(\hat{\theta}))}{m(s',F_{\alpha,x,\hat{x}}(\theta)) }. 
\end{align}
Since  $-m(s',F_{\alpha,x,\hat{x}}(\theta))$ is log-spm in $(s',  \alpha,\theta)$ for all $s'<s_0$, we obtain
\begin{align}\label{ineq-ratio2}
\liminf_{s'\uparrow s_0}\frac{-m(s',F_{\alpha,x,\hat{x}}(\hat{\theta}))}{-m(s',F_{\alpha,x,\hat{x}}(\theta)) }\geq \frac{-m(s,F_{\alpha,x,\hat{x}}(\hat{\theta}))}{-m(s,F_{\alpha,x,\hat{x}}(\theta))}. 
\end{align}
Combining inequalities \eqref{ineq-ratio1} and \eqref{ineq-ratio2} yields
\begin{align*}
-\frac{m(s,F_{\alpha,x,\hat{x}}(\theta))}{m(\hat{s},F_{\hat{\alpha},x,\hat{x}}(\theta))}\geq -\frac{m(s,F_{\alpha,x,\hat{x}}(\hat{\theta}))}{m(\hat{s},F_{\hat{\alpha},x,\hat{x}}(\hat{\theta}))}. 
\end{align*}
Thus, condition (ii) in  Lemma \ref{lm-tsc}  is satisfied. Similarly, since  $m(s',F_{\alpha,x,\hat{x}}(\theta))$ is log-spm in $(s',\alpha, \theta)$ for all $s'\geq s_0$, we obtain
\begin{align}\label{ineq-ratio3}
\frac{m(\hat{s},F_{\hat{\alpha},x,\hat{x}}(\theta))}{m(\hat{s},F_{\alpha,x,\hat{x}}(\theta))}\geq \limsup_{s'\downarrow s_0}\frac{m(s',F_{\hat{\alpha},x,\hat{x}}(\theta))}{m(s',F_{\alpha,x,\hat{x}}(\theta)) }. 
\end{align}
Since  $-m(s',F_{\alpha,x,\hat{x}}(\theta))$ is log-spm in $(s',  \alpha,\theta)$ for all $s'<s_0$, we obtain
\begin{align}\label{ineq-ratio4}
\liminf_{s'\uparrow s_0}\frac{-m(s',F_{\hat{\alpha},x,\hat{x}}(\theta))}{-m(s',F_{\alpha,x,\hat{x}}(\theta)) }\geq \frac{-m(s,F_{\hat{\alpha},x,\hat{x}}(\theta))}{-m(s,F_{\alpha,x,\hat{x}}(\theta))}. 
\end{align}
Combining inequalities \eqref{ineq-ratio3} and \eqref{ineq-ratio4} yields
\begin{align*}
-\frac{m(s,F_{\alpha,x,\hat{x}}(\theta))}{m(\hat{s},F_{\alpha,x,\hat{x}}(\theta))}\geq -\frac{m(s,F_{\hat{\alpha},x,\hat{x}}(\theta))}{m(\hat{s},F_{\hat{\alpha},x,\hat{x}}(\theta))}. 
\end{align*}
Thus, condition (iv) in  Lemma \ref{lm-tsc}  is satisfied. The result then follows directly from Lemma \ref{lm-tsc}. 
\end{proof}

\begin{proof}[Proof of Proposition \ref{prop-machina}]
Fix $\hat{x}>x$ and $\hat{\theta}>\theta$.  It follows from \eqref{eq-bothcase} that 
\begin{align*}
&U(\hat{x},\hat{\theta})-U(\hat{x},\theta)-[U(x,\hat{\theta})-U(x,\theta)]\\
=&\int_0^1 \int (v(s,\hat{x}, \alpha)-v(s,x, \alpha))d(F(s;\hat{\theta})-F(s;\theta))d\alpha. 
\end{align*}
By Assumptions (2) and (3), Lemma \ref{lm-tspm} implies 
\begin{align*}
U(\hat{x},\hat{\theta})-U(\hat{x},\theta)-[U(x,\hat{\theta})-U(x,\theta)]\geq 0,
\end{align*}
as desired. 
\end{proof}

%[\textbf{Proof for Proposition \ref{prop-case1-w}}]Lemma 5 in \citet{athey} implies $U_1(x,\theta)$ is SC1 in $\theta$. 
  
%Since $f(s;x)$ is  weakly SC2 in $(x;s)$, we have 
%\begin{align*}
%f(s,\hat{x})-f(s;x)\geq 0 \quad\Rightarrow \quad f(\hat{s},\hat{x})-f(\hat{s},x)\geq 0\quad\forall \hat{s}>s, \hat{x}>x.
%\end{align*}
%This implies 
%\begin{align*}
%\lim_{\hat{x}\rightarrow x^+} \frac{f(s,\hat{x})-f(s;x)}{\hat{x}-x}\geq 0 \quad\Rightarrow \quad \lim_{\hat{x}\rightarrow x^+} \frac{f(\hat{s},\hat{x})-f(\hat{s},x)}{\hat{x}-x}\geq 0\quad\forall \hat{s}>s.
%\end{align*}
%Thus, $f_x(s,x)$ is SC1. Since $v(s,\theta, x)$ is log-spm in $(s,\theta)$, 

%Fix $\hat{x}>x$. Applying integration by parts to \eqref{U-v} and taking limits yield
%\begin{align*}
%U_1(x,\theta)=-\int v_s(s,\theta, x)F_x(s,x)ds.  
%\end{align*}
%By Lemma 5 in \citet{athey},  Assumption \ref{as-case2-w} implies $U_1(x,\theta)$ is SC1.

\begin{proof}[Proof of Proposition \ref{prop-asset1-frechet}]
By Hadamard differentiability, 
\begin{align*}
U_1(x,\theta)=&\int  u_1(wr+x(s-r), F(\cdot; x,\theta)) (s-r) g(s;\theta)ds,
\end{align*} 
where $F(\cdot;x,\theta)$ is the distribution of the outcome $wr+x(s-r)$. Since $u_1(z,F)>0$  and  the expected return of $s$ under $\theta$ is larger than $r$,  this implies $U_1(0,\theta)>0$. The optimal investment $x>0$. Thus, the outcome $\kappa(s,x)= wr + x (s-r)$ increases in $s$ and is SC2 in $(x,s)$. Thus, Assumption (3) in Proposition \ref{prop-sc_1} is satisfied. Then the desired result follows directly from Proposition \ref{prop-sc_1} and Lemma \ref{lm-IDO}. 
\end{proof}

\section{Monotone Comparative Statics for Ambiguity}\label{sec-ambiguity}

In this section, we study MCS for ambiguity preferences, focusing on those that are Hadamard differentiable.  We believe the natural interpretation in these setting is that actions affect the outcomes and parameters affect ambiguity preferences since ambiguity preferences are seen as something exogenous they should not be impacted by actions.  

%When the agent perceives ambiguity, the agent's subjective beliefs are part of the ambiguity preferences. We thus focus on the environment in which 

To work with ambiguity preferences, we need to redefine our domain. Let $S$ denote the state space. Let $\kappa(s,x)$ be the outcome when state $s$ is realized and action $x$ is chosen. Assume $\kappa(s,x)$ is  increasing in $s$ and differentiable.  Let $u^*$ be the Bernoulli utility function and assume $u^*$ is differentiable, strictly increasing, and strictly concave. Let $\gamma(s;x):=u^*(\kappa(s,x))$, which can be seen as the act of the agent which is a mapping from the state space $S$ to utilities $\mathbb{R}$, given each $x$.  Let $\Gamma:=\{\gamma(\cdot;x)|x\in X\}$, endowed with the supnorm, be the set of feasible acts and the agent has preference on $\Gamma$. We assume that the parameter $\theta$ affects the agent's ambiguity preferences and let $I(\cdot;\theta):\Gamma\rightarrow \mathbb{R}$ be the agent's preference functional. Therefore, the agent's objective function is  $U(x,\theta)=I(\gamma(\cdot;x);\theta)$. 

Since the preference functional $I$ varies with models of ambiguity, we  consider three particular models of ambiguity: variational preferences,  smooth ambiguity preferences, and multiplier preferences. 

\subsection{Variational Preferences}
We start with the variational preferences of \citet{massimovar}:
\begin{align*}
I(\gamma(\cdot;x);\theta)=\min_{ F\in \Delta (S)}\Big( \int_{S} \gamma(s;x) d F(s)+ c(F,\theta)\Big),
\end{align*}
where $c(F,\theta)$ is a cost function parameterized by $\theta$. Intuitively, the agent's utility from action $x$ is obtained by minimizing her expected utility over the set of all probability distributions $\Delta (S)$ taking the cost $c(F,\theta)$ of each distribution $F$ into account. 

\begin{prop}\label{prop-spm-var}
Assume
\begin{enumerate}
    \item $c(F,\theta)$ is strictly convex in $F$ for each $\theta$;
    \item   $-c(F,\theta)$ is supermodular in $F$ for each $\theta$ and is supermodular in $(F, \theta)$; 
    \item  $\gamma(s; x)$ is supermodular in $(s,x)$.
\end{enumerate}
Then $U(x,\theta)$ is supermodular in $(x,\theta)$.
\end{prop}

We now explain the assumptions in the proposition. The next lemma shows that Assumption (1) ensures the preference functional is Hadamard differentiable.\footnote{\citet[Proposition 9]{quah2024} derived a similar MCS result without assuming strict convexity of $c$.}  

\begin{lemma}\label{lm-hadamard-var}
If Assumption (1) in Proposition \ref{prop-spm-var} holds, then $I$ is Hadamard differentiable.\footnote{See Appendix \ref{app-definition} for the definition for Hadamard differentiability on Banach spaces.} 
\end{lemma}

Fix $x$. \citet[Theorem 18]{massimovar} show that
\begin{align*}
U_{1}(x,\theta)=\int_{S} \gamma_{2}(s;x)dF^{*}(s,\theta), 
\end{align*}
where $\{F^{*}(\cdot,\theta)\}=\argmin_{F\in \Delta(S)} \Big( \int_{S} \gamma(s;x) d F+ c(F,\theta)\Big)$. The role of Assumptions (2) and (3) is given in the following lemma.
\begin{lemma}\label{lm-DD}
Fix $x$. If Assumptions (2) and (3) in Proposition \ref{prop-spm-var} hold, then  $F^{*}(\cdot,\hat{\theta})\succsim_{FOSD} F^{*}(\cdot,\theta)$ for all $\hat{\theta}>\theta$. 
\end{lemma}
Combining Lemma~\ref{lm-DD} with Lemma \ref{lm-tspm} yields $U_{1}(x,\theta)$ is increasing in $\theta$, which further implies $U$ is supermodular. That is,  Proposition~\ref{prop-spm-var} holds.

\subsection{Smooth Ambiguity Preferences}\label{sec-ambiguity-smooth}

We next study MCS under the smooth ambiguity model of \citet{kmm}. In the smooth ambiguity model, the agent considers the distribution of states is ambiguous in the sense that it is sensitive to some parameter $\lambda\in [0,1]$ whose true value is unknown. The ambiguity is characterized by a set $\Lambda=\{F(\cdot;\lambda)|\lambda\in [0,1]\}$ of plausible cdfs. The agent has a second-order belief over the set of priors $\Lambda$. For each plausible probability distribution $F(\cdot;\lambda)$, the agent computes the expected utility from action $x$ given by $\int_S \gamma(s;x)dF(s;\lambda)$.

We consider two ways in which $\theta$ can affect the agent's ambiguity preferences. One is $\theta$ affects the agent's second-order belief, which we denote by $\Pi(\lambda;\theta)$. Then the agent's utility is given by 
\begin{align*}
I(\gamma(\cdot;x),\theta)=\int_0^1 \phi\Big(\int_S \gamma(s;x)dF(s;\lambda)\Big)d\Pi(\lambda; \theta), 
\end{align*}
where $\phi$ is concave, differentiable, and strictly increasing. The function $\phi$ captures the agent's ambiguity attitude. Since $\gamma$ and $\phi$ are differentiable, the functional $I$ is Hadamard differentiable with
\begin{align}\label{eq-I}
U_{1}(x,\theta)=\int_{S}  u(s,\theta, \gamma(\cdot;x)) \gamma_2(s;x)ds,
\end{align}
and local utility function 
\begin{align}\label{eq-variation}
u(s,\theta,\gamma(\cdot;x))=\int_0^1 \phi_1\Big(\int_S \gamma(t;x)dF(t;\lambda)\Big)f(s;\lambda)\pi(\lambda; \theta)d\lambda, 
\end{align}
where $f(s;\lambda)$ and $\pi(\lambda;\theta)$ are the pdfs of $F(s;\lambda)$ and $\Pi(\lambda; \theta)$ respectively.

We now state our MCS result for smooth ambiguity. 

\begin{co}\label{co-ambiguity-general}
Assume 
\begin{enumerate}
    \item  $\kappa(s,x)$ has SC2 in $(x;s)$ a.e.;
    \item  $f(s;\lambda)$ is log-spm in $(s,\lambda)$ a.e.;
    \item  $\pi(\lambda;\theta)$ is log-spm in $(\lambda,\theta)$ a.e. 
\end{enumerate}
Then $U_1(x,\theta)$ is SC1 in $\theta$. 
\end{co}
By definition, $\gamma(s;x)=u^*(\kappa(s,x))$. Since $u^*$ is a strictly increasing function of $s$, Assumption (1) ensures that $\gamma(s;x)$ is SC2 in $(x;s)$ a.e. Assumption  (2) imposes an order on the set of possible beliefs: $F(\cdot;\hat{\lambda})$ dominates $F(\cdot;\lambda)$ in terms of the likelihood ratio for all $\hat{\lambda}>\lambda$.\footnote{$F(\cdot;\hat{\lambda})$ dominates $F(\cdot;\lambda)$ in terms of the likelihood ratio if $\frac{f(s;\hat{\lambda})}{f(s;\lambda)}$ is increasing in $s$. } Assumption (3) implies that a higher $\theta$ makes higher $\lambda$ relatively more likely. Intuitively, Assumptions (2) and (3) together imply that the agent with a higher $\theta$ considers higher states relatively more likely.

The other case in which the parameter can affect the preferences is $\theta$ affects the agent's ambiguity attitude. Slightly abusing notation, let $\Pi(\lambda)$ denote the agent's second-order belief with pdf $\pi(\lambda)$.  Then the agent's utility is given by 
\begin{align*}
\overline{I}(\gamma(\cdot;x),\theta)=\int_0^1 \overline{\phi}\Big(\int_S \gamma(s;x)dF(s;\lambda),\theta\Big)d\Pi(\lambda), 
\end{align*}
where $\overline{\phi}$ is concave, differentiable, and strictly increasing in the first argument for all $\theta$. The agent's objective function is thus $U(x,\theta)=\overline{I}(\gamma(\cdot;x),\theta)$ and
\begin{align}\label{eq-smooth-wealth}
U_{1}(x,\theta)=\int_{S}  \overline{u}(s,\theta, \gamma(\cdot;x)) \gamma_2(s;x)ds,
\end{align}
with local utility function 
\begin{align*}
\overline{u}(s,\theta,\gamma(\cdot;x))=\int_0^1 \overline{\phi}_1\Big(\int_S \gamma(t;x)dF(t;\lambda),\theta\Big)f(s,\lambda)\pi(\lambda)d\lambda. 
\end{align*}

\begin{co}\label{co-ambiguity-2}
Suppose Assumptions (1) and (2) from Corollary \ref{co-ambiguity-general} hold. Assume further that $\overline{\phi}_1(z,\theta)$ is log-spm in $(z,\theta)$ a.e. Then $U_1(x,\theta)$ is SC1 in $\theta$. 
\end{co}

The log-spm of $\overline{\phi}_1$ is equivalent to $-\frac{\overline{\phi}_{11}(z,\theta)}{\overline{\phi}_1(z,\theta)}$ decreasing in $\theta$. Thus, $\theta$ captures the agent's attitude towards ambiguity and a higher $\theta$ means less ambiguity aversion in the sense of   \citet[Corollary 3]{kmm}.

\subsection{Multiplier Preferences}
An important class of ambiguity preferences is the multiplier preferences (see, e.g., \citet{robust} and \citet{strmult}) which is in the intersection of variational preferences and smooth preferences. The multiplier preferences representation is given by
\begin{align*}
\min_{ F\in \Delta (S)}\Big( \int_{S} \gamma(s;x) d F(s)+ \mu R(F|| F^*(\cdot;\theta))\Big),
\end{align*}
where $F^*(\cdot;\theta)$ is a reference distribution, $R(F|| F^*(\cdot;\theta)):=\int_S \ln \frac{d F}{dF^*(\cdot;\theta)}dF$ is the relative entropy, and $\mu>0$ is a parameter that captures ambiguity aversion. This representation suggests the following interpretation: the reference distribution $F^*(\cdot;\theta)$ is the agent's ``best guess'' of the true probability law, but she is concerned that the true law may differ from $F^*(\cdot;\theta)$. To accommodate this concern with model misspecification, when evaluating each act, she takes all probability distributions into account, weighing more heavily those that are close to her best guess as measured by relative entropy.

As multiplier preferences are a special case of variational preferences and smooth preferences, the results  in the last two sections naturally extend to this context. Nonetheless, owing to specific functional forms associated with multiplier preferences, we can achieve stronger MCS results by directly analyzing them. In particular, as we focus on Savage acts, multiplier preferences are ordinally equivalent to the following EU preference functional\footnote{See, for example, \citet{strmult}.}
\begin{align}\label{multiplier}
U(x,\theta)=-\int_S \text{exp}\Big(-\frac{1}{\mu}  \gamma(s;x)\Big) dF^*(s;\theta). 
\end{align}
We focus on the case in which $\theta$ affects the agent's ambiguity preferences by affecting the reference distribution. In particular, we assume $F^*(\cdot;\theta)$ increases in $\theta$ with respect to the likelihood ratio.

Note that the objective function in \eqref{multiplier} can be viewed as a specific instance of the case studied in Appendix \ref{sec-MCS-u} in which $\theta$ affects the distribution while $x$ affects the outcome. The next result is an immediate consequence of  Proposition \ref{prop-sc}. 

\begin{co}\label{co-multiplier}
Assume
\begin{enumerate}
    \item $\kappa(s,x)$ has SC2 in $(x;s)$ a.e.;
    \item $f^*(s;\theta)$ is log-spm in $(s,\theta)$ a.e., where  $f^*(\cdot;\theta)$ is the pdf of $F^*(\cdot;\theta)$.\footnote{\citet{quah2024} show that this condition implies the cost function $c(F,\theta)=\mu R(F|| F^*(\cdot;\theta))$ satisfies Assumption (2) in Proposition \ref{prop-spm-var}.}
\end{enumerate}
Then $U(x,\theta)$ is SC2 in $(x;\theta)$. 
\end{co}
More generally, since  multiplier preferences conform to EU on the domain of Savage acts, we can derive a range of MCS by leveraging the results in Section \ref{sec-MCS-dist} and Appendix \ref{sec-MCS-u}, depending on how $x$ and $\theta$ affect the objective function.  Details are omitted.

\subsection{Application: Comparative Smooth Ambiguity}\label{app-smooth}

We use our approach in Appendix \ref{sec-ambiguity-smooth} to replicate the MCS result of ambiguity aversion in \citet[Corollary 1]{gollier2011} that less ambiguity aversion would increase demand for the uncertain asset.
 
We use the smooth ambiguity model of \citet{kmm}. Consider the following portfolio choice problem. An agent with wealth $w$  must decide how to invest it. She has two choices: a safe asset that pays a return $r>0$ and an uncertain asset that pays a nonnegative random return $s\in S$. If the agent spends $x$ on the uncertain asset and invests the remaining $w-x$ in the safe asset, she will end up with $xs + (w-x)r$. The agent considers the set of distributions $\Lambda=\{F(\cdot;\lambda)|\lambda\in [0,1]\}$ of $\tilde{s}$ possible. The agent's  second-order belief over the set $\Lambda$ is given by $\Pi$. We assume that the equity premium $\int_0^1 \int_S sdF(s;\lambda)d\pi(\lambda)>r$. Then by Lemma 1 in \citet{gollier2011}, the optimal investment in the uncertain asset is positive. %We thus only need to consider  $x>0$. 

We assume that the parameter $\theta$ affects the ambiguity attitude of the agent. Using the general setup in Appendix \ref{sec-ambiguity}, the outcome function is $\kappa(s;x)= wr + x (s-r)$, which is SC2 in $(x;s)$. Thus, Assumption (1) in Corollary \ref{co-ambiguity-general} is automatically satisfied. Since $u^*$ and $\overline{\phi}$ are strictly concave, straightforward verification shows that the objective function is quasiconcave in $x$.  Applying Corollary \ref{co-ambiguity-2} yields the following result that says the less ambiguity averse agent invests more in the uncertain asset.

\begin{co}\label{app-ambiguity}
Suppose $f(s;\lambda)$ is log-spm in $(s,\lambda)$ a.e. and $\overline{\phi}_1(z,\theta)$ is log-spm in $(z,\theta)$ a.e. Then an increase in $\theta$ leads to an increase in the optimal investment $x$ in the uncertain asset. 
\end{co}
Compared to \citet{gollier2011}, our approach illustrates that log-spm links the seemingly unrelated conditions on the set of distribution functions and the agent's ambiguity attitude.

\subsection{Application: Smooth Ambiguity and Wealth Effects}

In this application, we derive sufficient conditions for decreasing absolute ambiguity aversion under smooth ambiguity preferences and our result replicates \citet[Proposition 7]{gollier2015}.

We retain the portfolio choice framework from the previous section, but now investigate when the optimal investment in the uncertain asset increases in wealth $w$.  Hence, the parameter in this application is the wealth $w$. The objective function is given by
\begin{align*}
U(x,w)=\int_0^1 \overline{\phi}\Big(\int_S u^*(wr + x (s-r))dF(s;\lambda)\Big)d\Pi(\lambda). 
\end{align*}

\begin{co}\label{co-ambiguity-wealth}
Suppose $f(s;\lambda)$ is log-spm in $(s,\lambda)$ a.e. and $-\frac{u^*_{11}(z)}{u^*_{1}(z)}$ and $-\frac{\overline{\phi}_{11}(z)}{\overline{\phi}_{1}(z)}$ are decreasing in $z$. Then an increase in wealth leads to an increase in the optimal $x$. 
\end{co}

\subsection{Basic Formal Definitions and Proofs}\label{app-definition}

We first provide  definitions for Gateaux, Hadamard, and Frechet Differentiability in a Banach space $\mathcal{B}$. 
\begin{definition}
A function $\varphi: \mathcal{B} \rightarrow \mathbb{R}$ is \textbf{Gateaux differentiable} if at each $b\in \mathcal{B}$, there exists a bounded linear operator $D\varphi[b, \cdot]: \mathcal{B}\rightarrow \mathbb{R}$ such that for each $h\in  \mathcal{B}$, 
\begin{align*}
D\varphi[b, h]=\lim_{t\rightarrow 0} \frac{\varphi (b+th)-\varphi(b)}{t}.
\end{align*}
The operator  $D\varphi[b, \cdot]$ is called the \textbf{Gateaux derivative} of $\varphi$ at $b$. 
\end{definition}
%If the limit exists uniformly in $h$, we say $\varphi$ is Frechet differentiable at $b$ and  $D\varphi[b, \cdot]$ is called the \textbf{Frechet derivative} of $\varphi$ at $b$. 

\begin{definition}
A function $\varphi: \mathcal{B} \rightarrow \mathbb{R}$ is \textbf{Hadamard differentiable} if at each $b\in \mathcal{B}$, there exists a continuous and linear operator $D\varphi[b, \cdot]: \mathcal{B}\rightarrow \mathbb{R}$ such that for each $h\in  \mathcal{B}$, 
\begin{align*}
D\varphi[b, h]=\lim_{h'\rightarrow h, t\rightarrow 0} \frac{\varphi (b+th')-\varphi(b)}{t}.
\end{align*}
The operator  $D\varphi[b, \cdot]$ is called the \textbf{Hadamard derivative} of $\varphi$ at $b$. 
\end{definition}

\begin{definition}
A function $\varphi: \mathcal{B} \rightarrow \mathbb{R}$ is \textbf{Frechet differentiable} if at each $b\in \mathcal{B}$, there exists a continuous and linear operator $D\varphi[b, \cdot]: \mathcal{B}\rightarrow \mathbb{R}$ such that for each $h\in  \mathcal{B}$, 
\begin{align*}
\varphi (b+h)-\varphi(b)-D\varphi[b, h]=o(||h||). 
\end{align*}
The operator  $D\varphi[b, \cdot]$ is called the \textbf{Frechet derivative} of $\varphi$ at $b$. 
\end{definition}

If $\varphi$ is Gateaux differentiable and $D\varphi[b, \cdot]$ is continuous in $b$, then $\varphi$ is also Frechet differentiable. (See \citet{diff} Proposition A3.)

\begin{proof}[Proof of Lemma \ref{lm-hadamard-var}]
By Theorem 18 in \citet{massimovar}, Assumption (1) implies that $I$ is Gateaux differentiable. By Lemma 28 in \citet{massimovar}, we know that $I$ is Lipschitz continuous. It follows from Proposition 3.5 in \citet{diff_concepts} that these two guarantee the Hadamard differentiability of $I$. 
\end{proof}

\begin{proof}[Proof of Lemma \ref{lm-DD}]
Note that $\Delta(S)$ is lattice with respect to the order of FOSD. Moreover, for any $F,G\in \Delta(S)$, we have $(F\wedge G) (s)=\min \{F(s), G(s)\}$ and $(F\vee G) (s)=\max \{F(s), G(s)\}$. Thus,
\begin{align*}
\int_{S} \gamma(s;x) d F+\int_{S} \gamma(s;x) d G=\int_{S} \gamma(s;x) d (F\wedge G)+\int_{S} \gamma(s;x) d (F\vee G)\quad \forall x\in X.
\end{align*}
Combining this observation with Assumption (2) in Proposition \ref{prop-spm-var} yields that the function $-\Big(\int_{S} \gamma(s;x) d F+ c(F,\theta)\Big)$ is supermodular in $F$ and is supermodular in $(F,\theta)$. By \citet{milgrom1994}, the minimizer $F^{*}(\cdot,\theta)$ increases in $\theta$, as desired.  
\end{proof}

\begin{proof}[Proof of Corollary \ref{co-ambiguity-general}]
Observe that \eqref{eq-I} is an analogue of \eqref{eq-ambiguity}. To invoke Proposition~\ref{prop-body-ext1}, it suffices to show that $\gamma(s;x)$ satisfies SC2  and that $u(s,\theta,\gamma(\cdot;x))$ is log-spm in $(s,\theta)$ a.e. for all $x$. As discussed in the main text, the SC2 of $\gamma(s;x)$ is implied by Assumption~(1). The log-spm of $u$ follows directly from \eqref{eq-variation}, the condition $\phi_1 > 0$, Assumptions~(2) and~(3), and Lemma~2 in \citet{athey}. Consequently, Assumption~(1) of Proposition~\ref{prop-body-ext1} is satisfied, and Corollary~\ref{co-ambiguity-general} follows immediately.
\end{proof}

\begin{proof}[Proof of Corollary \ref{co-ambiguity-2}]
By Assumptin (1), $\gamma_2(s;x)$ is SC1 in $s$. For $U_1$ to be SC1 in $\theta$, we thus only need to show $u(s,\theta,\gamma(\cdot;x))$ is log-spm in $(s,\theta)$ a.e. for all $x$. Fix $x$.  By assumption, $u^*$ is an increasing function of $t$. Assumption (2) implies that $F(\cdot;\hat{\lambda})\succsim_{FOSD} F(\cdot;\lambda)$ for all $\hat{\lambda}>\lambda$. Thus, $\int_S u^*(\kappa(t,x))dF(t;\lambda)$ is an increasing function of $\lambda$. Then by log-spm of $\overline{\phi}_1$, we obtain that $\overline{\phi}_1\Big(\int_S u^*(wr+x(t-r))dF(t;\lambda),\theta\Big)$ is log-spm in $(\lambda,\theta)$ a.e. Combining this observation with Assumption (2) yields that $\overline{\phi}_1\Big(\int_S u^*(wr+x(t-r))dF(t;\lambda),\theta\Big)f(s,\lambda)$ is log-spm in $(s,\theta,\lambda)$ a.e. By Lemma 2 in \citet{athey}, $u(s,\theta,\gamma(\cdot;x))$ is log-spm in $(s,\theta)$ a.e.
\end{proof}

\begin{proof}[Proof of Corollary \ref{co-multiplier}]
Note that by \eqref{multiplier}, the multiplier preferences are equivalent to an EU functional form with local utility function $-\text{exp}\Big(-\frac{1}{\mu}  \gamma(s;x)\Big)$. Thus, Assumptions (1) and (4) in  Proposition \ref{prop-sc} are trivially satisfied. Combining with Assumptions (1) and (2) in Corollary \ref{co-multiplier},  Proposition \ref{prop-sc} implies that $U$ is SC2. 
\end{proof}
\begin{proof}[Proof of Corollary \ref{co-ambiguity-wealth}]
Since $u^*$ and $\overline{\phi}$ are strictly concave, straightforward verification shows that $U(x,w)$ is quasiconcave in $x$. Observe that \eqref{eq-smooth-wealth} reduces to 
\begin{align}\label{app-wealth}
U_{1}(x,w)=\int_{S}  \overline{u}(s,w, \gamma(\cdot;x)) u^*_{1}(wr + x (s-r))(s-r)ds
\end{align}
where
\begin{align*}
\overline{u}(s,w,\gamma(\cdot;x))=\int_0^1 \overline{\phi}_1\Big(\int_S u^*(wr + x (t-r))dF(t;\lambda)\Big)f(s;\lambda)\pi(\lambda)d\lambda. 
\end{align*}
Fix $x$. Since $-\frac{u^*_{11}(z)}{u^*_{1}(z)}$ is decreasing in $z$, it implies that $u^*_{1}(wr + x (s-r))$ is log-spm in $(s,w)$. Observe that $s-r$ satisfies SC1 in $s$. Thus, $U_1$ is SC1 in $w$ if $\overline{u}(s,w,\gamma(\cdot;x))$ is log-spm in $(s,w)$  for all $x$. To prove this, let
\begin{align*}
\varphi(w,\lambda):=\overline{\phi}_1\Big(\int_S u^*(wr + x (t-r))dF(t;\lambda)\Big).
\end{align*}
Note that $\varphi(w,\lambda)$ being log-spm  is equivalent to
\begin{align}\label{ratio}
\Big(-\frac{\overline{\phi}_{11}\Big(\int_S u^*(wr + x (t-r))dF(t;\lambda)\Big)}{\overline{\phi}_1\Big(\int_S u^*(wr + x (t-r))dF(t;\lambda)\Big)}\Big)\int_S u^*_{1}(wr + x (t-r))rdF(t;\lambda)
\end{align}
decreasing in $\lambda$. Note that since $\int_0^1 \int_S sdF(s;\lambda)d\pi(\lambda)>r$, Lemma 1 in \citet{gollier2011} shows that the optimal investment in the uncertain asset is positive. Thus, $u^*$ is an increasing function of $t$. Since $f$ is log-spm,  we have $F(\cdot;\hat{\lambda})\succsim_{FOSD} F(\cdot;\lambda)$ for all $\hat{\lambda}>\lambda$. Thus, $\int_S u^*(wr + x (t-r))dF(t;\lambda)$ is an increasing function of $\lambda$. Combining this observation with the assumption that  $-\frac{\overline{\phi}_{11}(z)}{\overline{\phi}_{1}(z)}$ is decreasing, we obtain that  the first term in \eqref{ratio} is decreasing in $\lambda$. Since $u^*$ is concave and $x>0$, $u^*_{1}$ is a decreasing function $t$. Also, since $F(\cdot;\lambda)$ increases in $\lambda$,  the second term in \eqref{ratio} is decreasing in $\lambda$. As a result, $\varphi(w,\lambda)$ is log-spm. Combining this observation with the log-spm of $f$,  yields that $\varphi(w,\lambda)f(s;\lambda)\pi(\lambda)$ is log-spm in $(s,w,\lambda)$. By Lemma 2 in \citet{athey}, $\overline{u}(s,w,\gamma(\cdot;x))$ is log-spm in $(s,w)$, as desired.
\end{proof}

%\section{When is $U(x,\theta)$ quasiconcave in $x$?} \label{sec-quasi}

%Note that $U(x,\theta)$ is quasiconcave in $x$ if and only if $-U_{1}(x,\theta)$ is SC1 in $x$. We can thus provide sufficient conditions on primitives for this to hold. For example, when the action $x$ affects the distribution (the case in Section \ref{sec-MCS-dist}), we can derive
%\begin{align}\label{eq-U1-dis}
%-U_{1}(x,\theta)=\int v_{1}(s,\theta, x) H_{2}(s; x)ds. 
%\end{align}
%Obviously, if $v_{1}(s,\theta, x) H_{2}(s; x)$ is increasing in $x$ for all $s$, then  $-U_{1}(x,\theta)$ is increasing in $x$ and, hence, SC1 in $x$. In the application in Section \ref{sec-machina4} under RDU, we obtain 
%\begin{align*}
%-U_{1}(x,\theta)=-\int \omega_{1}(F_{\tilde{s}}(s), \theta)u^*_{1}(wr+x(s-r),\theta) (s-r)f_{\tilde{s}}(s)ds.
%\end{align*}
%If $u^*_{11}<0$, then $-u^*_{1}(wr+x(s-r),\theta) (s-r)$ is increasing in $x$. As a result, $-U_{1}(x,\theta)$ is increasing and SC1 in  $x$.

%More generally, given \eqref{eq-U1-dis}, we can provide other sufficient conditions on the primitives such that $-U_{1}(x,\theta)$ is SC1 in $x$. For example, by applying Lemma \ref{lm-tsc}, another set of sufficient conditions is $v_{1}>0$, $H_{2}(s; x)$ is SC1 in $(-s,x)$, and $\frac{v_{1}(s,\theta, x)}{v_{1}(\hat{s},\theta, x)}$ and $\frac{H_{2}(\hat{s}; x)}{H_{2}(s; x)}$ are both increasing in $x$ for all $\hat{s}>s$.\footnote{We assume  $H_{2}(s; x)$ is SC1 in $-s$ to be consistent with Assumption (2) in Proposition \ref{prop-case1}.}

 \end{document}